\documentclass[3p,times]{elsarticle}
\journal{Journal of Fluids and Structures}

\usepackage[hidelinks]{hyperref}
\usepackage{amssymb}
\usepackage{booktabs,adjustbox} % for tables
\usepackage{siunitx}
\usepackage[UKenglish]{isodate}
\usepackage{graphicx}
\usepackage[fleqn]{mathtools} % includes amsmath and aligns equations to the left
\usepackage{xcolor}
\usepackage[UKenglish]{isodate}
\cleanlookdateon
\usepackage{caption,subcaption}
\usepackage{numcompress}\biboptions{authoryear}

\makeatletter
\def\ps@pprintTitle{%
   \let\@oddhead\@empty
   \let\@evenhead\@empty
   \let\@oddfoot\@empty
   \let\@evenfoot\@oddfoot
}
\makeatother

\begin{document}

\begin{frontmatter}
\title{Evolution of an eroding cylinder in single and lattice arrangements}
\author[ouraddress]{James N. Hewett\corref{mycorrespondingauthor}}
\ead{james@hewett.nz}
\cortext[mycorrespondingauthor]{Corresponding author}
\author[ouraddress]{Mathieu Sellier}
\address[ouraddress]{Department of Mechanical Engineering, University of Canterbury, Christchurch 8140, New Zealand}

\begin{abstract}
The coupled evolution of an eroding cylinder immersed in a fluid within the subcritical Reynolds range is explored with scale resolving simulations. Erosion of the cylinder is driven by fluid shear stress. K\'arm\'an vortex shedding features in the wake and these oscillations occur on a significantly smaller time scale compared to the slowly eroding cylinder boundary. Temporal and spatial averaging across the cylinder span allows mean wall statistics such as wall shear to be evaluated; with geometry evolving in 2-D and the flow field simulated in 3-D. The cylinder develops into a rounded triangular body with uniform wall shear stress which is in agreement with existing theory and experiments. We introduce a node shuffle algorithm to reposition nodes around the cylinder boundary with a uniform distribution such that the mesh quality is preserved under high boundary deformation. A cylinder is then modelled within an infinite array of other cylinders by simulating a repeating unit cell and their profile evolution is studied. A similar terminal form is discovered for large cylinder spacings with consistent flow conditions and an intermediate profile was found with a closely packed lattice before reaching the common terminal form.
\end{abstract}
\end{frontmatter}

\section{Introduction}
Erosion due to flowing fluid occurs in a wide range of contexts. Wind erosion sculpt rocks, forming natural arches and other shapes dependent on the surrounding topology and rock properties. Yardangs are streamlined erosional wind forms \citep{Ward1979,Ward1984} whereas hoodoos are columns, pillars and toadstool rock forms \citep{Scheidegger1958,Wang2005}. These eroded rock formations affect the surrounding wind patterns which consequently influence the rock formations \citep{Ward1984,Scheidegger1958}. Another example of this coupled effect between a fluid and solid body is river meanders \citep{Leopold1960}; the naturally occurring curved paths of rivers. Simulations of streams with an initially straight channel have developed into these meandering patterns \citep{Howard1984}. Erosion also features in biology with blood flows through arteries where plaque erosion and plaque rupture can be fatal \citep{Farb1996}. Ruptures often occur in regions of high wall shear stress $\tau_w$ upstream of the plaque \citep{Groen2007}. In this paper we examine the time evolution of an initially circular cylinder with both laminar and turbulent upstream conditions in high speed unidirectional flow (Reynolds number $\text{Re}=\num{27000}$) using numerical simulations. We find that the cylinder tends to a terminal form and then erodes self-similarly as found in experiments \citep{Ristroph2012}; dependent on the upstream conditions but independent of the initial or transitional shape.

Two modelling approaches exist for simulating the interface between phases: Eulerian (captures the boundary) and Lagrangian (tracks the boundary). The Eulerian approach is based on the immersed boundary method where the fluid and structure interaction (FSI) is modelled with the fictitious domains method and has been applied for cylinders in Stokes flow ($\text{Re} \ll 1$, quasi steady flow) \citep{Golay2011}. However, this approach has difficulty in accurately modelling the fluid properties at the interface such as $\tau_w$. The Lagrangian approach involves exclusively simulating the fluid phase and remeshing the computational domain as the interface deforms. This remeshing allows the standard meshing procedures for resolving near-wall velocity gradients to accurately capture surface features such as $\tau_w$: either directly resolving the viscous sublayer or using wall treatment models. Recently, \citet{Mercier2014} have simulated erosion of soil from a turbulent jet using a 2-D axisymmetric model. They treated the flow as steady and used the $k-\epsilon$ and $k-\omega$ Reynolds averaged Navier-Stokes (RANS) turbulence models. However, remeshing is a computationally expensive task and their run times were one month with a cluster of 8 CPUs. We have used the Lagrangian approach because $\tau_w$ was selected as the driving mechanism of erosion and a highly resolved mesh near the boundary was sought to accurately simulate $\tau_w$ at this moderate $\text{Re}$.

We simulated the erosion of a circular cylinder in the subcritical flow regime at $\text{Re} = \num{27000}$ with two types of configurations with particular attention to the fluid structure coupling. First, we simulated a single cylinder to validate our model against theory \citep{Moore2013} assuming a laminar approximation and then against an experiment \citep{Ristroph2012} using scale resolving simulations. Second, we investigated a lattice of cylinders which are sparsely separated and then a closely packed lattice where the flow transitions from shear layer reattachment (closely packed) to vortex shedding (smaller eroded cylinders with large spacings). There are no previous studies on simulating the erosion of a cylinder with unsteady turbulence in literature to the best of our knowledge, and nor dynamic meshing with a boundary deforming over such a significant change in curvature and scale.

\section{Methods}
\subsection{Geometry and flow conditions}
Flow over a cylinder was simulated with the same conditions as an experiment by \citet{Ristroph2012} to validate our model against, and a cylinder within a lattice of cylinders was also modelled. The cylinder eroded as a function of the local wall shear stress $\tau_w$. The cylinder had an initial radius, $a_0 = \SI{18}{\milli\meter}$, giving a Reynolds number, $\text{Re} = 2 u_\infty a_0 / \nu = \num{27000}$ where $u_\infty = \SI{0.61}{\meter\per\second}$ is the freestream velocity and $\nu = \SI{8e-7}{\meter\squared\per\second}$ the kinematic viscosity of water. This $\text{Re} = \num{27000}$ is within the subcritical flow regime for a cylinder in cross flow where the wake is completely turbulent and there is a laminar boundary layer separation point on the top and bottom of the cylinder \citep{Sumer2006}. $\text{Re}$ scales linearly with $a$ and typically reduces by a factor of four in our simulations and remains in this subcritical flow regime.

The Strouhal number is defined as
\begin{equation}
\text{St} = \frac{2 f_v a}{u_\infty}
\label{eqn:Strouhal_number}
\end{equation}
where $f_v$ is the vortex shedding frequency and is the inverse of the vortex period $T_v = 1/f_v$. The effective cylinder radius is time-dependent $a = a(t)$, reduces as the cylinder erodes and $2a$ is defined as the width of the cylinder (normal to flow). This characteristic length scale $2a$ controls the flow field behaviour including shedding vortex properties and Reynolds number.

Based on Prandtl boundary layer theory, the cross sectional area $A$ of the eroding body follows a $4/3$ power law in time $t$ \citep{Moore2013} with
\begin{equation} \label{eqn:power_law}
A(t) \sim A_0 \left(1-\frac{t}{t_\text{end}}\right)^{4/3}
\end{equation}
where $A_0$ is the initial area and $t_\text{end}$ the vanishing time of the body. The simulations were unable to reach $t_\text{end}$ (where no material remained). Instead, a simulation-dependent final time $t_f$ is defined where $t_f < t_\text{end}$. The shear stress can be estimated as
\begin{equation} \label{eqn:tau_star}
\tau^* = \rho \sqrt{\frac{\nu u_\infty^3}{a}}
\end{equation}
where $\rho = \SI{998.2}{\kilogram\per\metre\cubed}$ is the density of the fluid (water, as per the experiment). The $\tau_0^*$ is a fixed characteristic stress and was used for non-dimensionalising.

Drag and lift coefficients were calculated as the total surface integral of the pressure and skin friction on the body. The projected areas, normal (for drag) and parallel (for lift) to the flow, are time-dependent and therefore these values were calculated for each mesh update.

The root mean square intensities of fluctuations in $\tau_w$ are greatest around the separation point \citep{Yokuda1990}. This separation point $\theta_\text{sep}$ is where the $\tau_w$ vanishes as the boundary layer separates from the wall \citep{Achenbach1968}.

\subsubsection{Single cylinder}
\begin{figure} \centering
  \begin{subfigure}[b]{0.5\textwidth} \centering
    \includegraphics{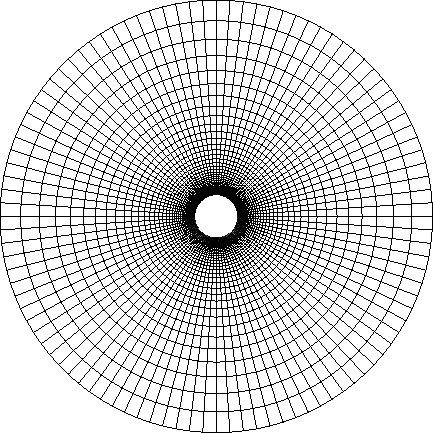}
    \caption{\centering\label{fig:mesh_single}}
  \end{subfigure}%
  \begin{subfigure}[b]{0.5\textwidth} \centering
    \includegraphics{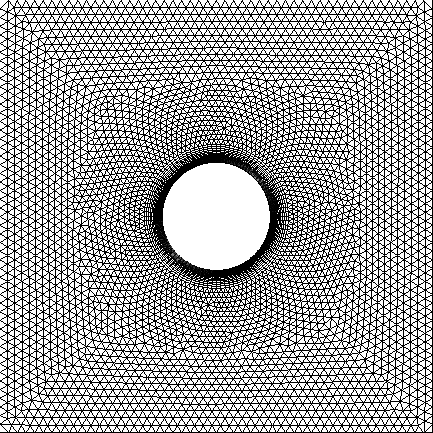}
    \caption{\centering\label{fig:mesh_lattice}}
  \end{subfigure}
  \caption{Side view of meshes with flow from left to right for the (\subref{fig:mesh_single}) single cylinder and (\subref{fig:mesh_lattice}) repeating unit cell configuration ($L/2a_0 = 4$). All cases had $a_0 = \SI{18}{\milli\metre}$ and a refined O-grid mesh around the cylinder.\label{fig:mesh_info}}
\end{figure}

A curvilinear O-type orthogonal grid was used with $50 \times 100$ (radial $\times$ circumferential) cells in the cross sectional plane as shown in Figure~\ref{fig:mesh_single}. Mesh nodes were clustered near the cylinder with $\Delta r / a_0 = \num{4e-3}$ for the first cell height giving $y_\text{max}^+ \approx 4$ throughout the simulations including the deformed mesh of the eroded cylinder. One spanwise cell was used to simulate the laminar 2-D case and 32 spanwise cells were used for the unsteady 3-D cases; this resolution of the spanwise flow features was indistinguishable compared with a 64 cell deep grid. The spanwise length was $8a_0$ which is adequate to accurately simulate the time-averaged statistics, as reasoned by \citet{Lysenko2014}. The O-grid had an outer radius of $10a_0$ for the single cylinder cases giving \num{5e3} cells in 2-D and \num{1.6e5} cells in 3-D.

A laminar, uniform inlet speed of $u_\infty$ was applied on the upstream side and an outflow boundary condition was applied on the downstream side to handle the vortices leaving the domain (non-constant pressure distributions across the vortices). Periodic boundaries were applied on the sides (parallel to flow) to allow vortices to translate across the span of the cylinder (unsteady turbulence does not have instantaneous symmetry). A no-slip wall was applied to the cylinder boundary.

\subsubsection{Lattice of cylinders} \label{sec:lattice_of_cylinders}
Three common types of cylinder arrangements are tandem (cylinders aligned in streamwise direction), side-by-side (perpendicular to flow) and staggered (often used in heat exchangers). Experiments, and more recently scale resolving simulations to capture the vortex shedding, have been undertaken with these simple configurations and few papers exist to compare results for the staggered case \citep{Sumner2010}. A lattice of cylinders is a structured array of cylinders which are equally spaced with longitudinal $L$ (parallel to flow) and transverse $T$ (normal to flow) centre to centre distances. This lattice is a combination of tandem and side-by-side configurations and we used a uniform grid with $T = L$.

\begin{figure} \centering
  \begin{subfigure}[b]{0.5\textwidth} \centering
    \includegraphics{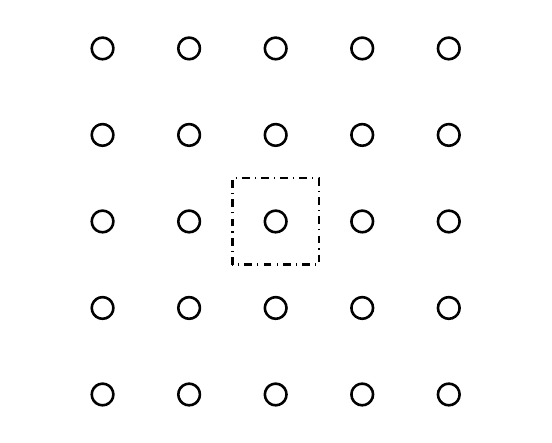}
    \caption{\centering\label{fig:unitcell_sparse}}
  \end{subfigure}%
  \begin{subfigure}[b]{0.5\textwidth} \centering
    \includegraphics{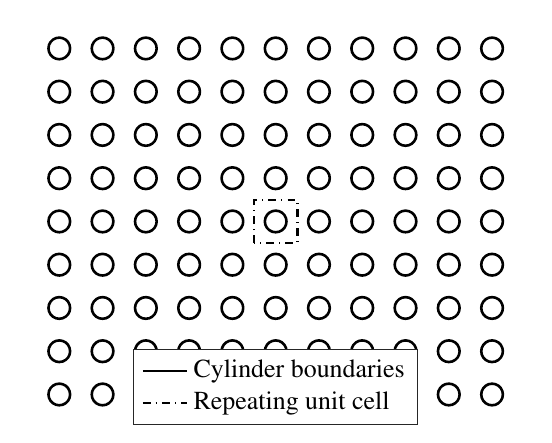}
    \caption{\centering\label{fig:unitcell_packed}}
  \end{subfigure}
  \caption{Lattice of cylinders are shown for both the (\subref{fig:unitcell_sparse}) sparsely spaced ($L/2a_0 = 8$) and (\subref{fig:unitcell_packed}) closely packed ($L/2a_0 = 4$) configurations with their respective repeating computational unit cells used in the simulations with periodic boundary conditions.\label{fig:unitcell_info}}
\end{figure}

The lattice of cylinders were modelled using a repeating unit cell of one cylinder; the two arrangements explored are shown in Figure~\ref{fig:unitcell_info}. This computational domain is significantly smaller than modelling a finite number of cylinders; for example, three tandem cylinders where the surroundings and especially the wake and near-wall regions of the cylinders need to be accurately resolved. The bounding box of the unit cell controls the spacings $T$ and $L$; altering these spacings gives disparate flow fields for equal $\text{Re}$. Two basic types of interferences contribute to the fluid behaviour \citep{Zdravkovich1987}, namely: wake interference (downstream cylinders are either partially or completely within the wake of the upstream cylinder) and proximity interference (cylinders are sufficiently close to affect one another, typically side-by-side, but neither is submerged in the wake of the other). Several flow patterns exist for two tandem cylinders \citep{Igarashi1981}, and a simplified classification scheme \citep{Sumner2010} can be described as:
\begin{itemize}
\item extended-body regime: small pitch ratios of $1 < L/l < 1.2 - 1.8$ behave as a single bluff body. The downstream cylinder lies within the vortex formation region of the upstream cylinder. The gap between the cylinders contain mostly stagnant fluid.
\item reattachment regime: intermediate pitch ratios of $1.2 - 1.8 < L/l < 3.4 - 3.8$ have shear layer reattachment between the cylinders. Eddies can form between the cylinders and gap dynamics widely vary.
\item co-shedding regime: large pitch ratios of $L/l > 3.4 - 3.8$ have vortex shedding between the cylinders. The downstream cylinder is outside the vortex formation region of the other cylinder and experiences periodic impacts from the shed vortices.
\end{itemize}
where $l$ is the maximum length of the cylinder in the streamwise direction ($l_0 = 2a_0$ and $l \approx 2a$ when the aspect ratio $2a/l$ is near unity). All pitch ratio ranges are approximate, depend on $\text{Re}$ and the above values are from \citet{Zdravkovich1987} with similar flow conditions to our case (subcritical flow regime). K\'arm\'an vortex shedding occurs downstream of both cylinders for all cylinder spacings presented in this paper.

We first modelled a sparsely spaced lattice analogous to the tandem configuration of two cylinders with $L/l_0 = 4$. Cylinders with a transverse spacing of $T/2a_0 = 4$ reside outside the proximity interference region \citep{Sumner2010} and therefore the upstream flow from each cylinder is dictated by the shedding vortices from the preceding cylinder. A critical pitch ratio $L/l = 4.05$ at $\text{Re} = \num{7000}$ and $L/l = 3.47$ at $\text{Re} = \num{27000}$ \citep{Ljungkrona1993} exists where the flow between the gap alternates between the reattachment and co-shedding regimes. This lattice avoids this bistable flow by having a spacing above the critical ratio throughout the erosion process; $\text{Re}$ decreases with the eroding cylinder ($\text{Re}_f \approx 7000$) but the pitch ratio increases at a similar rate ($L/l_f \approx 16$). Our second lattice configuration had a closer packing with $L/l_0 = 2$ which transitioned from the complex reattachment regime, with proximity interference, to the co-shedding regime without proximity interference as the cylinder eroded.

A similar O-type grid was used around the cylinder with $20 \times 100$ cells with an outer radius of $4 a_0 / 3$. The accuracy of this case was retained from the validation case by preserving the first cell height, spanwise length and number of spanwise cells. The surrounding block was filled with unstructured tetrahedral cells yielding a total of \num{2.8e5} and \num{1.2e5} cells for the large spacing and closely packed cases respectively. The mesh was recreated twice during the course of the simulation for the close packed case: first at $t/t_f = 0.68$ (two thirds through the simulation) and again at $t/t_f = 0.95$ (near the end of the simulation but not $t_\text{end}$). The remeshing procedure was employed to maintain a similar grid resolution throughout the simulation with a significant increase in fluid volume ratio (no additional cells were normally created as described in Section~\ref{sec:dynamic_meshing}). The O-grid had the same thickness and radial distribution of cells to retain the $y_\text{max}^+$ values.

Periodic boundary conditions were applied to all sides of the bounding cuboid and a specified mass flow rate was applied in the streamwise direction to attain the correct $\text{Re}$ (using 5 sub-iterations to determine the pressure gradient for the desired mass flow rate). A phase lag of fluctuating lift forces between two tandem cylinders exists at $\text{Re} = \num{65000}$ \citep{Alam2003}. However we have assumed a zero phase lag by forcing synchronised vortex shedding across the lattice using a repeating unit cell. A no-slip wall was applied to the cylinder boundary.

\subsection{Numerical methods}
ANSYS Fluent R17.0 was used as the solver and MATLAB R2016a for data analysis. Simulations were run on a standard desktop computer with an Intel Core i7-6700 CPU. Fluent was run in serial mode for the 2-D cases and in parallel mode for the 3-D cases. Production runs with unsteady 3-D simulations of a single cylinder took $20$ hours to complete and $65$ hours for the lattice cases.

The iterative time advancement scheme with the SIMPLEC segregated solver was used with under-relaxation factors set at unity to improve iterative convergence. Absolute residuals of $\num{1e-4}$ were set as the convergence criteria for each time step; giving $\approx 25$ inner iterations. The bounded second order implicit transient formulation was employed with an initial time step $\Delta \tilde{t}_0 = \SI{50}{\milli\second}$. Second order for pressure and third order MUSCL for momentum, turbulent kinetic energy and specific dissipation rate was used for the spatial discretisation. High order schemes and the iterative (instead of non-iterative) time advancement scheme were chosen as these settings yielded the best results with the coarse time step and grid.

\subsection{Viscous models}
The flow was first approximated with a steady laminar viscous model in 2-D. Shear in the wake was ignored in accordance with \citet{Moore2013} by setting $\tau_w = 0$ where the circumferential angle from the stagnation point $\theta > \theta_{\text{sep}}$ for calculating the eroding normal speed $v_n$. The flow field upstream of the cylinder is laminar, however, unsteady turbulence occurs in the wake downstream. This assumption of neglecting the erosion in the wake is reasonable for an approximation, because this region had the lowest erosion rate observed in experiments \citep{Ristroph2012}.

Scale resolving turbulence models, such as large eddy simulations (LES), can resolve the turbulent structures in the wake behind the cylinder. Few studies exist using LES for the flow over a cylinder within the subcritical flow regime. \citet{Lysenko2014} have recently investigated using various subgrid scale models at $\text{Re} = \num{20000}$ and compared their results to experiments and simulations. They used a mesh size of $440 \times 440 \times 32$ (radial $\times$ circumferential $\times$ spanwise), a total of $\num{1.24e7}$ cell volumes, taking a cluster of 256 cores in parallel approximately 60 hours for each run. In theory, a slightly more refined mesh would be required at our higher $\text{Re} = \num{27000}$ to capture the boundary layer with the same resolution; because the boundary layer thickness \citep{Schlichting1999} behaves like $\delta \approx r/\sqrt{Re}$ where $r$ is the radial coordinate. Furthermore, the simulation time step $\Delta \tilde{t}$ would need to be refined to have a matching Courant number.

This issue of long computational times for accurate simulations is also apparent for arrays of cylinders. The wake, and subsequently the upstream region due to the periodicity, should have a fine grid to capture the vortex structures and accurately simulate $\tau_w$. \citet{Uzun2012} have simulated two tandem cylinders with detached eddy simulations (DES) at a critical spacing of $L/l = 3.7$ and $\text{Re} = \num{166000}$ giving a bistable flow regime which alternates between reattachment and vortex shedding. Their finest mesh had 133 million grid points taking 23 days using 816 cores in parallel. Primary vortex periods were highly resolved with 2899 time steps; whereas our simulations had seven time steps per vortex period. However, their coarser grids yield similar results and our time and mesh resolution was chosen as the coarsest settings while still adequately resolving $\tau_w$.

The unsteady flow field changes over time due to the fluid-structure interaction with the eroding cylinder. Flow features, and consequently the shear forces acting on the cylinder, constantly change over the evolution of the erosion process. Developing the flow field and extracting the relevant flow characteristics at every discrete mesh update, considering the above computational times, would be impractical. Instead, we have applied a pragmatic approach whereby the distribution of $\tau_w$ (the key parameter that influenced the erosion rate) was first verified against the case of a rigid (non-deforming) circular cylinder in cross-flow in the subcritical flow regime. Second, the evolution of the cylinder profile was validated against the experimental results of \citet{Ristroph2012}. The mesh was considered adequate if $\tau_w$ was verified (with particular attention to the separation point $\theta_{\text{sep}}$) and the cylinder evolution matched the experiment. An independence study on both the temporal and spatial discretisation was carried out and the $\tau_w$ profile of the converged grid is compared with that of the coarse mesh and experimental data in Figure~\ref{fig:DES_wallshear}.

DES are composed of LES and RANS models to help mitigate the challenge of impractical computational costs for high Reynolds number separated flows \citep{Spalart1997}. The separation point is sensitive to the accuracy of modelling the transition from laminar to turbulence; shielded DES and delayed DES \citep{Menter2004,Spalart2006} employ a DES scale limiter which is solution dependent (not only grid dependent). We used the improved delayed DES turbulence model \citep{Shur2008}, designed for simulating wall boundary layers at moderate $\text{Re}$ with LES, with the SST $k-\omega$ RANS model in Fluent.

\subsection{Erosion}
The cylinder from the experiment \citep{Ristroph2012} eroded and was observed for $t_f = \SI{115}{\minute}$ and reduces to approximately a third of its initial width, with an estimated vanishing time of $t_\text{end} = 140 \pm \SI{2}{\minute}$. The cylinder boundary eroded at a normal speed of $v_n \approx \SI{1}{\centi\meter\per\hour}$ which is very slow compared to the freestream velocity, giving a ratio of $v_n/u_\infty \approx \num{5e-6}$. Therefore the movement of the receding cylinder wall has an insignificant effect on the momentum of the surrounding flow.

The eroded material would act as sort of a sand blaster downstream of the cylinders and have an effect in the lattice arrangements. These new particles may alter the flow properties and influence the separation of the laminar boundary layers and the vortex shedding. We assumed the eroded material had a negligible impact on the flow physics because of its low volume fraction, considering the small ratio of $v_n/u_\infty \approx \num{5e-6}$.

Clay was used as the material for the cylinder in the experiment and erodes such that $v_n \propto |\tau_w|$ \citep{Ristroph2012}, the wall shear stress magnitude; from now $|\tau_w|$ is written as $\tau_w$ for simplification. The dimensionless normal erosion velocity can be written based on the formulation of \citet{Moore2013} as
\begin{equation} \label{eqn:erosion_velocity}
\textbf{v}_n = \left( \frac{\tau_w}{\tau_0^*} + \epsilon \sqrt{\frac{L}{L_0}} \frac{(\kappa - \kappa_\text{mean})}{\kappa_0} \right) \hat{\textbf{n}}
\end{equation}
where $\epsilon$ is the curvature factor, $L = \sum s_i$ the total arc length of all arc segments $s_i$, $\kappa$ the local curvature and $\hat{\textbf{n}}$ the unit normal outwards from the fluid domain. The first term on the right hand side is erosion from shear stress $v_\tau$ and the second term is from the curvature correction $v_\kappa$.

Erosion of the cylinder was driven by $v_\tau$, whereas $v_\kappa$ was used to smooth the sharp edges (for example at stagnation points where $\tau_w = \SI{0}{\pascal}$) with an effective weighting of $\epsilon$. The $\sqrt{L}$ factor was used to scale the influence of smoothing with $\tau_w$ as the body vanishes as predicted by Equation~\ref{eqn:tau_star}. The mean curvature $\kappa_\text{mean}$ was subtracted from $\kappa$ to conserve the mass such that material loss was only due to the action of shear stress \citep{Moore2013}. These variables are divided by initial values $L_0$ and $\kappa_0$ to obtain the non-dimensionalised Equation~\ref{eqn:erosion_velocity}.

\citet{Menger1930} introduced a definition of curvature for three discrete points by fitting a circle through these points. This Menger curvature was used to calculate the local curvature at nodes ($\textbf{x}_i$) about its nearest two neighbours ($\textbf{x}_{i-1}$ and $\textbf{x}_{i+1}$) in the circumferential direction with
\begin{equation}
\begin{split}
& \kappa (\textbf{x}_{i-1}, \textbf{x}_i, \textbf{x}_{i+1}) = \\
& \frac{\sqrt{(\textbf{x}_{i-1} \textbf{x}_i + \textbf{x}_i \textbf{x}_{i+1} + \textbf{x}_{i-1} \textbf{x}_{i+1}) (\textbf{x}_{i-1} \textbf{x}_i + \textbf{x}_i \textbf{x}_{i+1} - \textbf{x}_{i-1} \textbf{x}_{i+1})(\textbf{x}_{i-1} \textbf{x}_i - \textbf{x}_i \textbf{x}_{i+1} + \textbf{x}_{i-1} \textbf{x}_{i+1})(-\textbf{x}_{i-1} \textbf{x}_i + \textbf{x}_i \textbf{x}_{i+1} + \textbf{x}_{i-1} \textbf{x}_{i+1})}}{\textbf{x}_{i-1} \textbf{x}_i \cdot \textbf{x}_i \textbf{x}_{i+1} \cdot \textbf{x}_{i-1} \textbf{x}_{i+1}}
\end{split}
\end{equation}

The nodes on the cylinder boundary were displaced on a separate time scale than the fluid with
\begin{equation} \label{eqn:displace_nodes}
\Delta \textbf{x} = \eta a_0 \textbf{v}_n
\end{equation}
where $\eta$ is the erosion factor and $|\textbf{v}_n|$ is of order unity. The inverse $1/\eta$ provides a rough estimate of the number of mesh updates $N_\text{end}$ required for the cylinder to vanish. The product of these variables $N_\text{end} \eta$ provides a quantitative metric to compare the rate of erosion for the cases.

\subsection{Spatial and temporal averaging}
Two distinct timescales were observed in the experiment: (1) the fluid with several vortex shedding cycles per second; (2) eroding boundary motion across a couple of hours. We assumed the unsteady 3-D flow features across the cylinder would average across the span of the cylinder after many vortex cycles; this dependence of only $\textbf{v}_n = \textbf{v}_n(\theta, t)$ was observed in the experiments with a 2-D shape evolution in the axial direction of the cylinder \citep{Ristroph2012}.

The 3-D computational domain was still required because the vortices have 3-D effects including unsynchronised vortex shedding across the cylinder causing out of phase hydrodynamic forces which are smaller when compared to a 2-D flow assumption. The $\tau_w$ was averaged across the span of the cylinder and a constant cross section was maintained (2-D shape evolution). Far fewer vortex periods were required to converge mean transient statistics compared to observing a single slice because of collecting this data across 32 spanwise cells instead of one cell; providing effectively as many vortex periods for a simulation time of $T_v$.

\begin{figure} \centering
  \includegraphics{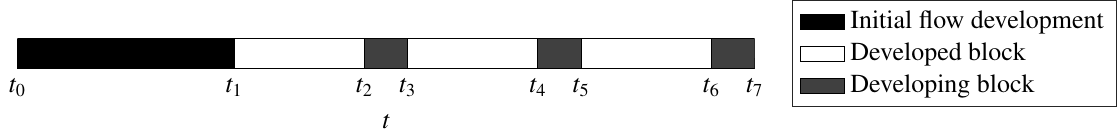}
  \caption{Repeating time periods of developed and developing blocks after the initial flow field has been established. Three sets of blocks are shown here; typical simulation runs had over \num{600} sets of developed and developing blocks.\label{fig:average_time}}
\end{figure}

The flow field was first initialised and developed without effects of erosion; shown as the first time block in Figure~\ref{fig:average_time} from $t_0$ to $t_1$. In practice, the developed 3-D flow was saved and then new runs were initialised with this solution; requiring fewer time steps per run for the first time block in subsequent simulations. The time-averaged wall shear stress of each boundary face was stored in memory and was reset at each new time block. Mesh updates occured at the end of each developed block $t_2,t_4,t_6...$ after the time-averaged $\tau_w$ was no longer changing over time. Developing blocks allowed the flow to redevelop after each mesh deformation; the solver was generally well-behaved where residuals spiked across only one time step. The dynamic mesh model in Fluent requires a first order time discretisation and this setting was returned to the bounded second order scheme on the time step following each mesh update.

\begin{figure} \centering
  \includegraphics{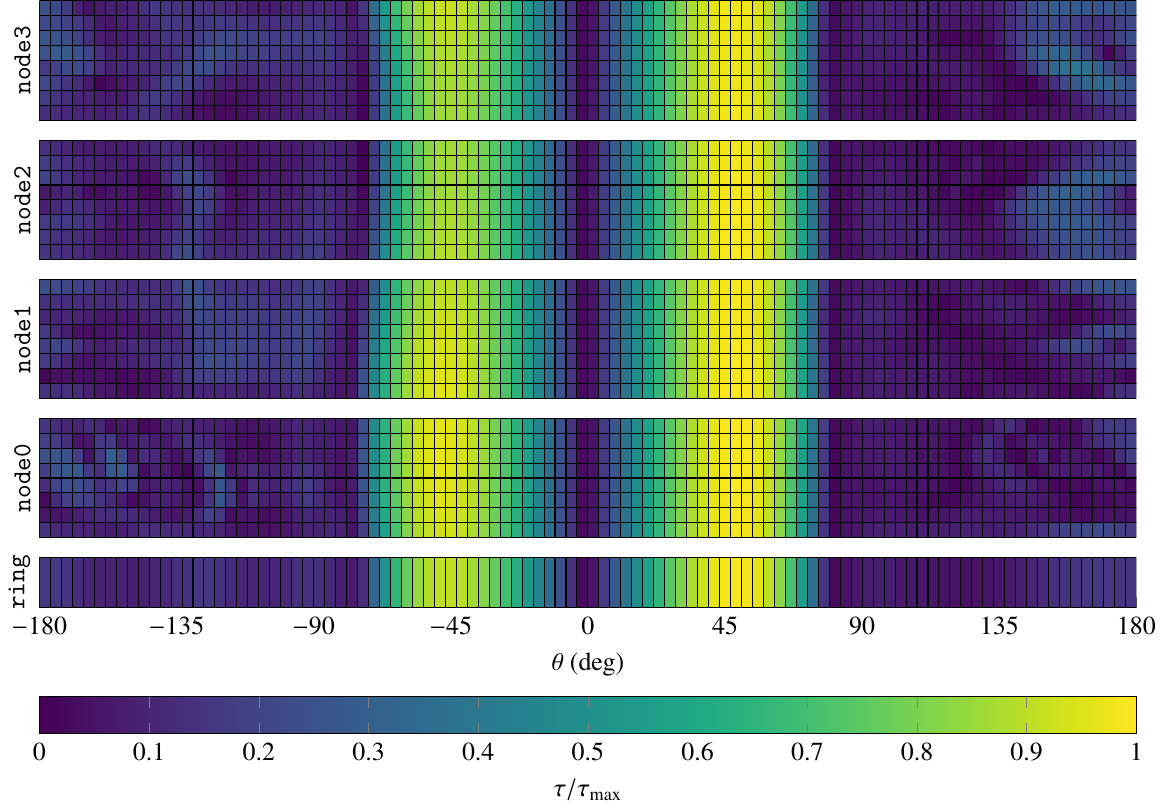}
  \caption{Wall shear stress distribution of the cylinder across compute nodes at an instance in time. The ring, $\texttt{averageRing}$, contains the mean of each angle across all spanwise boundary faces.\label{fig:average_shear}}
\end{figure}

An array variable, $\texttt{averageRing}$, contained averaged values for each face angle of the structured grid as shown in Figure~\ref{fig:average_shear}. Each compute node had its own $\texttt{averageRing}$, and $\texttt{node0}$ received and averaged across all faces using message passing macros. A $\texttt{primaryRing}$ array of the first ring on $\texttt{node0}$ enabled monitoring of a slice of the cylinder. Lastly, the $\texttt{nodeRing}$ array contained the locations and details of all nodes on the cylinder wall boundary: made possible by assuming that the geometry morphs only in 2-D. Partitions were aligned parallel with the flow to reduce the influence of interpartition approximations (as only one finite volume cell overlapped between partitions).

\subsection{Dynamic meshing} \label{sec:dynamic_meshing}
\begin{figure} \centering
  \begin{subfigure}[b]{0.5\textwidth} \centering
    \includegraphics{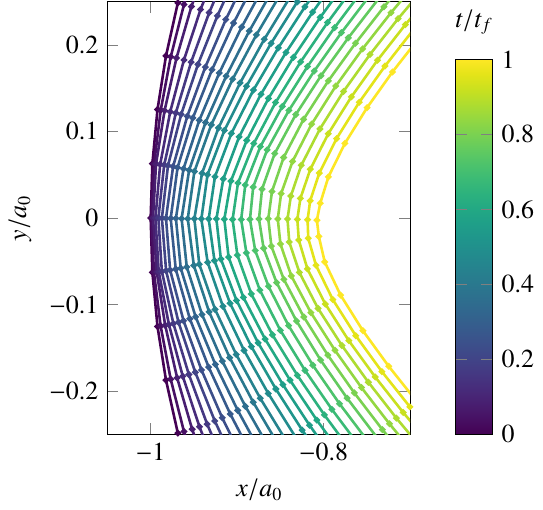}
    \caption{\centering\label{fig:nodeshuffle_without}}
  \end{subfigure}%
  \begin{subfigure}[b]{0.5\textwidth} \centering
    \includegraphics{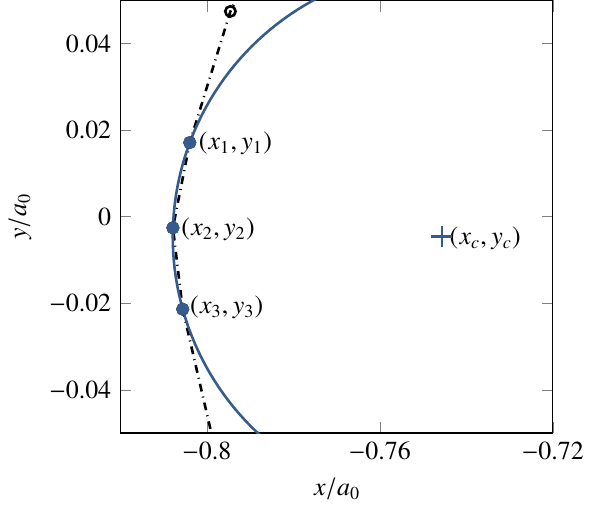}
    \caption{\centering\label{fig:nodeshuffle_diagram}}
  \end{subfigure}
  \caption{(\subref{fig:nodeshuffle_without}) Evolution of the cylinder boundary without the node shuffle algorithm. Nodes have clustered near the stagnation point due to the high curvature change. (\subref{fig:nodeshuffle_diagram}) Close-up featuring the node triplet nearest to the stagnation point at $t=t_f$. The active shuffle node $(x_2,y_2)$ is surrounded by its left $(x_3,y_3)$ and right $(x_1,y_1)$ neighbours. The circle intersects all three nodes and has its centroid at $(x_c,y_c)$.\label{fig:nodeshuffle_intro}}
\end{figure}

The mesh naturally bunched and eventually caused overlapping nodes in the initial trial simulations as shown in Figure~\ref{fig:nodeshuffle_without}; particularly near the stagnation point. Forcing the node trajectories along constant $\theta$ with the geometric centroid improved mesh quality; but the mesh distribution became non-uniform ($\theta_i$ was uniform although $s_i$ was not). Ultimately a node shuffle algorithm, detailed in Section~\ref{sec:node_shuffle}, was designed to elegantly deform the cylinder boundary nodes without losing mesh quality.

Diffusion-based smoothing was chosen as the method for displacing the interior nodes using the dynamic mesh model in Fluent. This mesh motion was governed by the diffusion equation
\begin{equation}
\nabla \cdot (\gamma \nabla \textbf{m}) = 0
\end{equation}
where $\gamma$ is the diffusion coefficient and $\textbf{m}$ the mesh displacement velocity. This equation was discretised using the standard finite volume method and the algebraic multigrid solver was used to iteratively solve the generated matrix. Node positions were then updated by interpolating this cell centred solution.

A boundary distance formulation was selected with $\gamma = 1/b^{\beta}$ where $b$ is the normalised boundary and $\beta$ is a parameter such that $\beta \geqslant 0$. Setting $\beta = 0$ gives $\gamma = 1$ such that the interior of the mesh diffuses uniformly, whereas a higher value of $\beta$ reduces the relative node displacement of the mesh near the moving boundary (and the far field absorbs the deformation). Therefore a relatively high $\beta = 2$ was chosen to preserve the grid resolution across the boundary layer: the critical region to accurately simulate the $\tau_w$ profile.

Fluent is a cell centred finite volume solver and therefore all field variables are stored at the centroid of each cell and face. These variables, including wall shear stress and pressure, were stored in $\texttt{averageRing}$ and $\texttt{primaryRing}$ whereas the geometric values, including profile coordinates and curvature, were stored in $\texttt{nodeRing}$. The boundary was deformed by displacing the nodes which required interpolation of the face values onto the nodes. A high order interpolation method was sought to capture the local $v_{\tau}$ component. Catmull-Rom splines \citep{Catmull1974} have $C^{1}$ continuity, local control and interpolation; and were employed to interpolate at nodes from the nearest four faces in the circumferential direction.

The faces and nodes in their respective arrays were ordered according to $\theta$ which was relative to the current geometric centroid. The centroid receded as the cylinder eroded and the profile deformed; the centroid position was calculated at each mesh update. $N$ nodes, with coordinates ${(x_i,y_i)}_{i=1}^N$, formed a closed polygon around the cylinder in the plane normal to the cylinder axis. The signed area of this polygon is
\begin{equation}
A = \frac{1}{2} \sum_{i=1}^N (x_i y_{i+1} - x_{i+1} y_i)
\end{equation}
and its centroid is given by
\begin{equation}
(x_\text{centroid},y_\text{centroid}) = \frac{1}{6A} \left( \sum_{i=1}^N (x_i + x_{i+1}) (x_i y_{i+1} - x_{i+1} y_i) , \sum_{i=1}^N (y_i + y_{i+1}) (x_i y_{i+1} - x_{i+1} y_i) \right)
\end{equation}

Oscillating ripples formed across nodes in regions of highest curvature and profile changes; a result of overcompensating curvature from the $v_{\kappa}$ term. One solution for this problem was reducing $\eta$ to have smaller node displacements at each mesh update, however this method linearly increases the computational effort. A second solution was to use an exponential smoothing factor in time
\begin{equation} \label{eqn:smoothing_factor}
\textbf{v}_{n} = \alpha \textbf{v}_{n,t} + (1-\alpha) \textbf{v}_{n,t-1}
\end{equation}
where $\alpha$ is the smoothing factor and $0 < \alpha < 1$. Applying $\alpha = 0.75$ dampened the fluctuating ripples while still allowing the profile evolution to respond to local transient dynamics. This solution was implemented with the same $\eta$ and thus no additional intermediate mesh updates were required, resulting in a stable solution without increasing the computational effort.

$\text{St}$ remains relatively uniform across a wide range of $\text{Re}$ in the subcritical flow regime \citep{Schewe1983}, especially within the Reynolds range of our simulations \citep{Son1969,Norberg2003}: reducing from $\text{Re} = \num{27000}$ to $\text{Re} \approx \num{7000}$ as the cylinder cross section reduces from $a=a_0$ to $a/a_0 \approx 0.25$ (both in the experiment and for our simulations). Similarly, $\text{St}$ is directly proportional to $a$ as shown in Equation~\ref{eqn:Strouhal_number}, therefore $T_v$ scales linearly with $a$. A \textit{transient} time step was used to retain the same resolution of the vortex oscillations throughout the cylinder evolution with 
\begin{equation} \label{eqn:time_step}
\Delta \tilde{t} = \frac{a}{a_0} \Delta \tilde{t}_0
\end{equation}
where $\Delta \tilde{t}_0$ is the initial simulation time step. This time step was modified after each mesh update and is separate from the time scale $t$ used for describing the erosion rate.

\subsection{Node shuffle algorithm} \label{sec:node_shuffle}
A uniform distribution of nodes around the cylinder boundary was used for the initial mesh and was desired throughout the simulation. However, nodes tended to bunch in regions of high curvature deformation as shown in Figure~\ref{fig:nodeshuffle_without}. This phenomenon  ultimately led to \textit{pinching} of the mesh where nodes overlapped causing negative (non-physical) cell volumes.

\begin{figure} \centering
  \begin{subfigure}[b]{0.5\textwidth} \centering
    \includegraphics{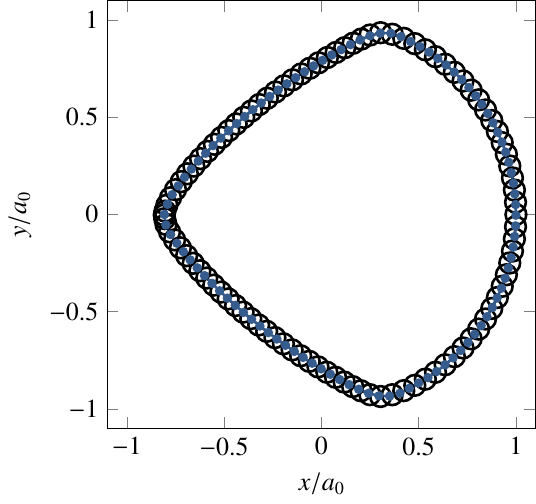}
    \caption{\centering\label{fig:nodeshuffle_compare}}
  \end{subfigure}%
  \begin{subfigure}[b]{0.5\textwidth} \centering
    \includegraphics{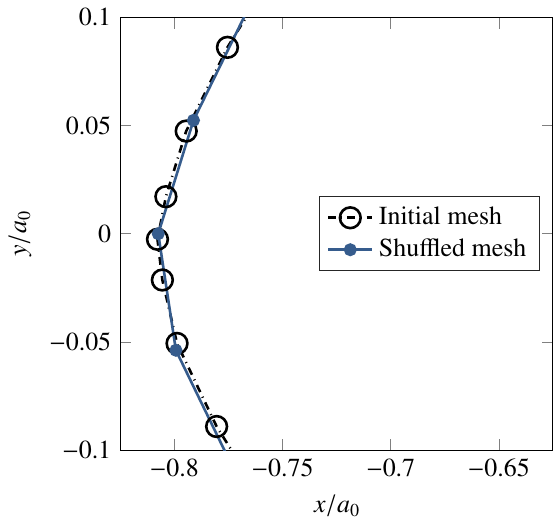}
    \caption{\centering\label{fig:nodeshuffle_comparezoom}}
  \end{subfigure}
  \caption{Comparison of the cylinder boundary before and after the node shuffling. (\subref{fig:nodeshuffle_compare}) The cylinder outline of both meshes shows that the profile was preserved throughout the shuffling process. (\subref{fig:nodeshuffle_comparezoom}) Close-up where the centre node is anchored at the stagnation point ($\theta = \ang{0}$). The shuffled mesh has uniform arc lengths between each node.\label{fig:nodeshuffle_comparison}}
\end{figure}

We have designed an algorithm to uniformly distribute the nodes around an arbitrary profile described by its nodes. One aim was to preserve the mesh quality as the cylinder evolved from the initial circle to its terminal form without reordering the nodes. Nodes are shuffled in an iterative fashion and the local curvature of each node is conserved (to avoid artificially altering the curvature). Subsequently, a second aim of preserving the global profile was achieved (Figure~\ref{fig:nodeshuffle_compare}). Lastly, a third aim was to position a node at the stagnation point to help capture the highest curvature which was found at the leading edge (Figure~\ref{fig:nodeshuffle_comparezoom}).

\begin{figure} \centering
  \includegraphics{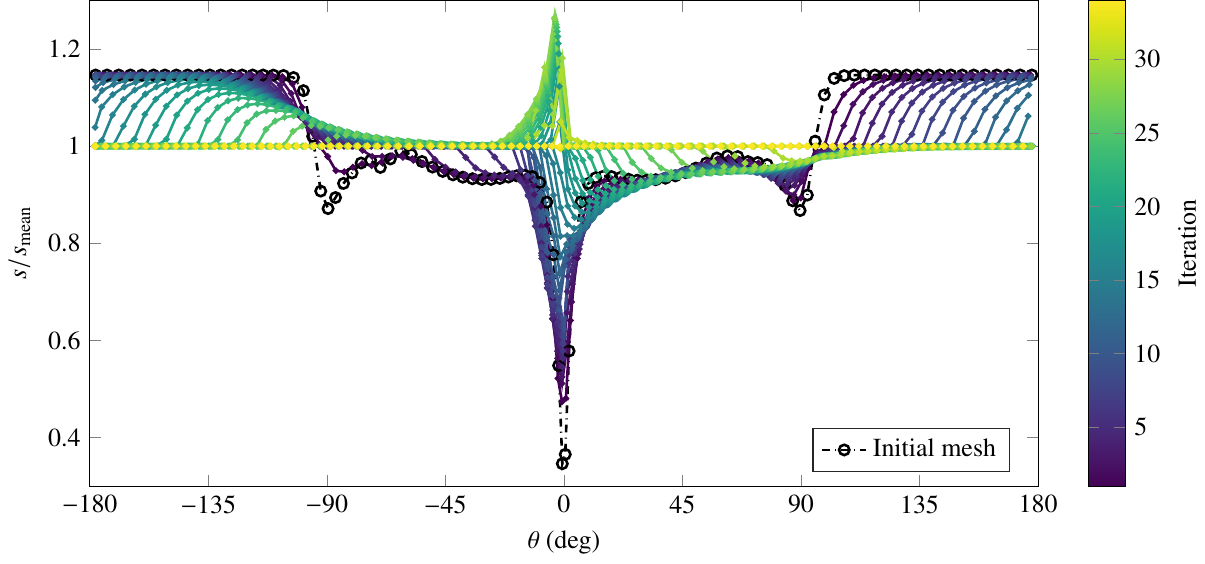}
  \caption{Normalised arc lengths around the cylinder starting from an initial mesh where the arc length ratio $s_\text{max}/s_\text{min} = 3.31$. The initial mesh was generated in the absence of the node shuffling algorithm. Each arc length $s$ gradually tended toward $s_\text{mean}$ as the nodes propagated around the cylinder. $\xi = 0.025$ yielded $s_\text{max}/s_\text{min} = 1$ after $34$ iterations with a shuffled perimeter error of $\SI{0.02}{\percent}$.
  \label{fig:nodeshuffle_arclength}}
\end{figure}

Each iteration cycles through all nodes starting at the lead node (nearest to the stagnation point) and continues in a clockwise direction; this process is illustrated with a video available in the supplementary material. A node triplet is defined for each node as well as a circle, with radius $r_c$, which intersects the nodes as shown in Figure~\ref{fig:nodeshuffle_diagram}. The active shuffle node moves either left or right along the circle arc to decrease or increase the left hand arc length ($s_{1-2}$). The target arc length is $s_\text{target} = s_\text{mean}$ except for the lead node which always tends to $\theta = \ang{0}$. The shuffled arc length is restricted along the circle arc such that
\begin{equation}
(0.5 - \xi) s_{1-3} \leqslant s_\text{shuffled} \leqslant (0.5 + \xi) s_{1-3}
\end{equation}
where $\xi$ is the shuffling factor ($0 < \xi < 0.5$) and $s_{1-3}$ the arc length between the neighbouring nodes. Nodes are shuffled slowly around the cylinder with a low value of $\xi$; yielding a greater accuracy when compared to using a high $\xi$. We used $\xi = 0.05$ in our simulations to ensure the cylinder profile was preserved accurately. Typically, two iterations were sufficient for converging the uniformity of the nodes at each mesh update. A lower $\xi$ was chosen for shuffling nodes around a poor mesh (Figure~\ref{fig:nodeshuffle_arclength}) to highlight the gradual movement of nodes around the cylinder boundary.

A key advantage of anchoring a node is that the cells do not rotate around the cylinder causing the mesh to skew. Anchoring the lead node at the stagnation point encourages a consistent (but evolving) shape as the leading edge propagates due to erosion. Having similar node angles on the cylinder boundary throughout mesh updates enhances the stability of the solver by reducing the change in field variables. Therefore, the computational domain did not require initialising after each mesh deformation.

Figure~\ref{fig:nodeshuffle_arclength} shows how the nodes are shuffled from left to right (direction of increasing $\theta$; clockwise) throughout the iterations. Cell faces with $s/s_\text{mean} \neq 1$ gradually tend to unity. The spike at the lead node is caused by the slack from the node shuffles during an iteration; the lead node is \textit{anchored} at $\theta = \ang{0}$ (after reaching this position).

\section{Results}
Four distinct cases are investigated in this paper with two viscous models, 2-D or 3-D, steady or unsteady and type of cylinder configuration. First, a 2-D steady laminar model of a single cylinder is analysed and compared with a Prandtl-based method. Next, a 3-D unsteady DES model of a single cylinder is validated against experimental data. Lastly, two cases are simulated with 3-D unsteady DES models and are of a sparsely spaced, and then closely packed, lattice of cylinders.

\subsection{Steady laminar approximation}
We started with simulating steady laminar flow to help verify our model. A weighting of $\epsilon = 0.1$ for the curvature component gave shape deformation changes similar to that found from the Prandtl-based method. The characteristic shear stress was $\tau_0^* = 3.17~\text{Pa}$.

\begin{figure} \centering
  \includegraphics{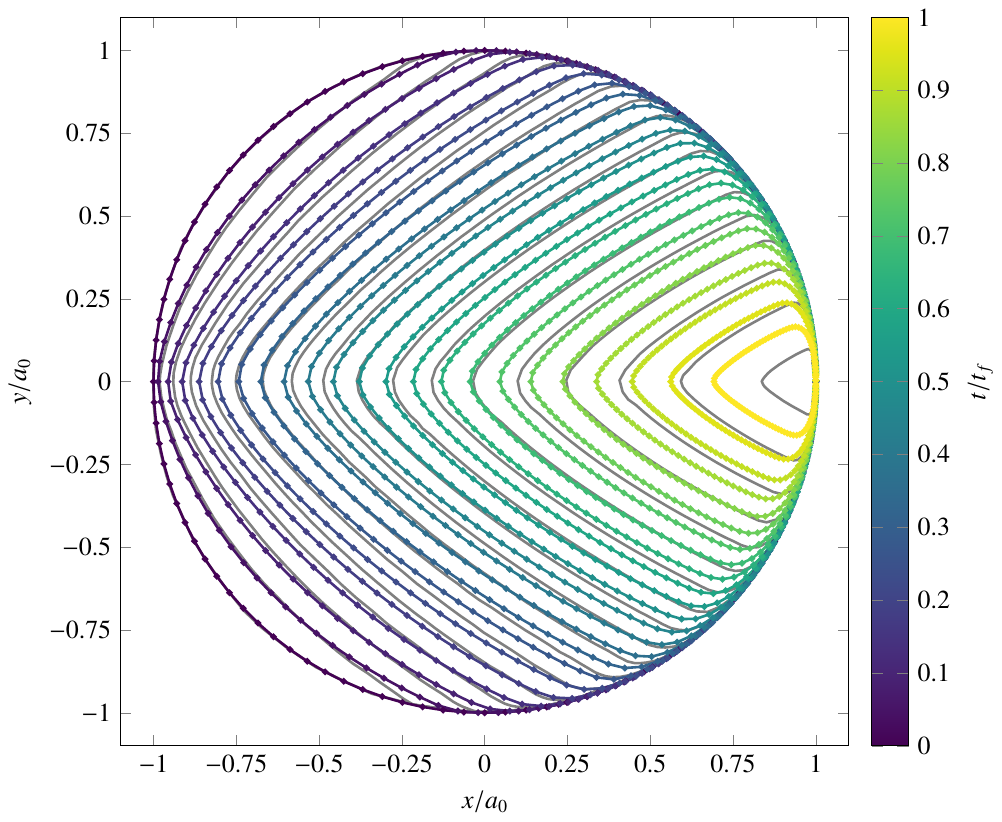}
  \caption{Erosion of a clay cylinder in water (flow is from left to right) assuming steady laminar conditions in 2-D. The coloured lines show the evolution of the cylinder profile where boundary nodes are represented with points. The erosion factor $\eta = \num{1.4e-3}$ and 1100 mesh updates were performed (every 50th is shown here). The solid grey lines are from a Prandtl-based method \citep{Moore2013} where the profiles are on a separate time scale, are linearly spaced in time and extend beyond $t_f$ by \SI{4.2}{\percent}.
  \label{fig:laminar_profile}}
\end{figure}

The cylinder eroded over time and approached a terminal form as shown in Figure~\ref{fig:laminar_profile} due to the changing velocity field. The cylinder eroded fastest at the areas of highest wall shear stress ($\theta = \pm \ang{45}$), tending towards a rounded triangular shape (pointed upstream) from $t/t_f \approx 0.3$. The cylinder then continued eroding at a similar rate across the upwind side causing a self-similar evolution which matched closely with the model \citep{Moore2013} and experiments \citep{Ristroph2012}.

The deforming mesh behaved well when using the shuffling algorithm and anchoring the node at the stagnation point as confirmed when comparing the grey (solution of \citet{Moore2013}) and coloured dotted contours (our simulation) in Figure~\ref{fig:laminar_profile}. By comparison, the mesh distorted and collapsed at $t/t_f = 0.75$ with anchoring a node at $\theta \approx -\ang{180}$ (in the wake region), and broke down at $t/t_f = 0.25$ without the shuffling algorithm entirely. Enhancing the robustness of the mesh deformation allowed us to simulate the erosion of the cylinder from the initial circular shape up to near its vanishing point without manually remeshing the geometry several times.

\begin{figure} \centering
  \includegraphics{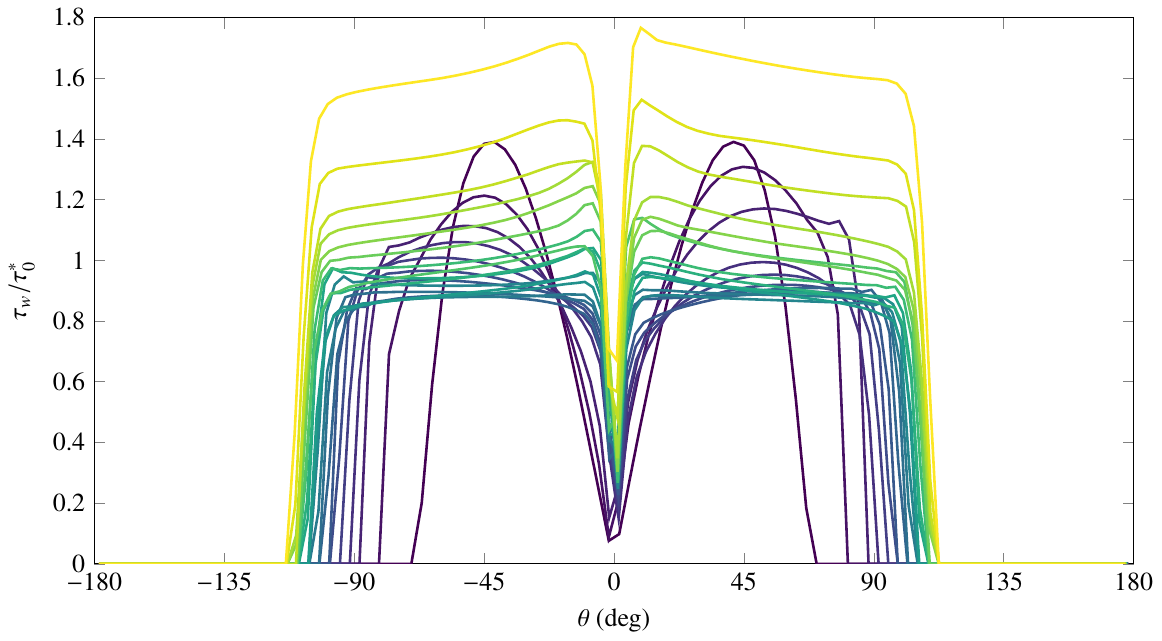}
  \caption{Wall shear stress distribution around the cylinder for the 2-D steady laminar case (same colour legend as Figure~\ref{fig:laminar_profile}). The shear force acting on the cylinder becomes uniform as the cylinder approaches its terminal form. Wall shear stress beyond separation was ignored in the moving mesh algorithm as suggested by \citet{Moore2013}.\label{fig:laminar_wallshear}}
\end{figure}

A uniform wall shear stress distribution on the upwind side of the cylinder was obtained as the cylinder reached its terminal form as shown in Figure~\ref{fig:laminar_wallshear}. The Falkner-Skan similarity solutions for flow over a wedge suggest that a right angled nose exhibits a uniform wall shear stress across the surface \citep{Moore2013}. The shear stress was symmetric about the stagnation point because the flow was treated as steady; vortices were not able to develop, which would in turn produce oscillatory forces. The shear force on the downstream side of the cylinder was ignored because a laminar viscous model was used which cannot capture the turbulent wake effects. This assumption was based on the model proposed by \citet{Moore2013} and reasoned with observing significantly lower erosion rates on the leeward face compared to the windward face in the experiment \citep{Ristroph2012}. Furthermore, the shear in the wake would be significantly overpredicted in 2-D as the spanwise motion of the shedding vortices would not be resolved \citep{Mansy1994,Williamson1995}. The curvature component of erosion $v_\kappa$ was modelled across the complete cylinder.

\begin{figure} \centering
  \begin{subfigure}[b]{0.5\textwidth} \centering
    \includegraphics{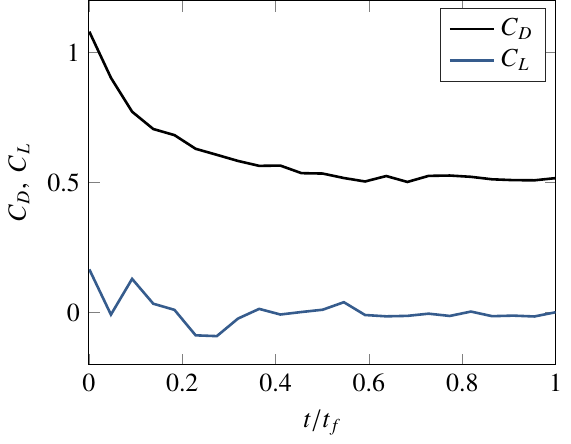}
    \caption{\centering\label{fig:laminar_coefficients}}
  \end{subfigure}%
  \begin{subfigure}[b]{0.5\textwidth} \centering
    \includegraphics{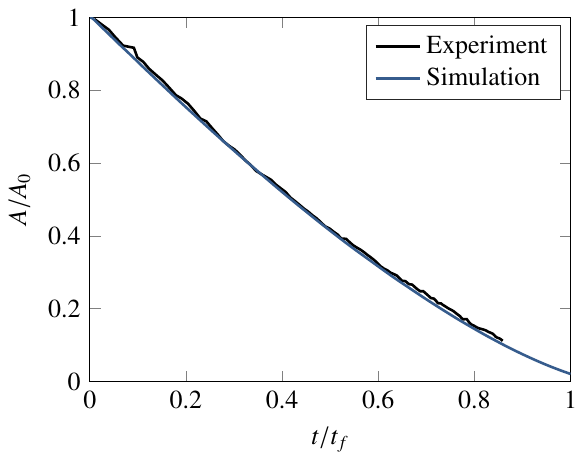}
    \caption{\centering\label{fig:laminar_area}}
  \end{subfigure}
  \caption{Single deforming cylinder using the 2-D steady laminar approximation. (\subref{fig:laminar_coefficients}) Drag and lift coefficients of the cylinder from an initial circular shape until its final eroded form. (\subref{fig:laminar_area}) Cross sectional area evolution normalised by the initial area. Both the experiment \citep{Ristroph2012} and simulation follow a $4/3$ power law.\label{fig:laminar_info}}
\end{figure}

The drag coefficient, using the time-dependent cross sectional area $A = A(t)$, initially reduced as the cylinder evolved from its circular shape to its more streamlined rounded triangular form, as shown in Figure~\ref{fig:laminar_coefficients}. After reaching its terminal form where it eroded in a self-similar manner, the drag remained constant with $C_D = 0.5$ towards the vanishing point. The lift remained near zero because the flow was modelled in steady state (forcing a symmetric flow field over the cylinder, with no oscillations). Cross sectional area reduced according to the $4/3$ power law in Equation~\ref{eqn:power_law} and matched closely with experimental data as shown in Figure~\ref{fig:laminar_area}.

\subsection{Single cylinder with vortex shedding}
The simulated flow field needed to accurately model the experimental conditions to predict the erosion; including the unsteady vortex shedding. The development of the oscillating vortex shedding and evolving wall shear stress distributions are visualised with videos in the supplementary material. Key flow features are first verified against both experiments and LES simulations from literature as shown in Table~\ref{tab:DES_summary}. These experiments and simulations are within the subcritical flow regime allowing a quantitative comparison using dimensionless values. \citet{West1993} observed a relatively wide range in $C_D$ and $C^{\prime}_L$ because the turbulence intensity $u^{\prime}/U$ was varied between $\SI{0.2}{\percent}$ and $\SI{7.5}{\percent}$; inlet conditions can play a significant role in the drag and lift forces on the cylinder.

\begin{table}[] \centering
%\footnotesize
\caption{Summary of experimental and LES studies of a circular cylinder in cross flow within the subcritical flow regime}
\label{tab:DES_summary}
\begin{tabular}{@{}llllllll@{}}
\toprule
Study                                      & Method   & $\text{Re}$         & $C_D$       & $C^{~\prime}_L$ & $\text{St}$    & $-C_{P,b}$  & $\theta_{\text{sep}}$ \\ \midrule
\citet{Lim2002}                            & EXP      & \num{20000}  & $1.2$       &                 & $0.187$ & $1.07$      &                \\
\citet{Son1969}                            & EXP      & \num{20000}  &             &                 &         &             & $\ang{83}$         \\
\citet{Norberg1994,Norberg2003}            & EXP      & \num{27000}  &             & $0.48$          & $0.192$ & $1.22$      &                \\
\citet{West1993}                           & EXP      & \num{27000}  & $1.1-1.4$   & $0.49-0.74$     &         &             &                \\
\citet{Schewe1983}                         & EXP      & \num{30000}  & $1.1$       & $0.31$          & $0.19$  &             &                \\
\citet{Son1969}                            & EXP      & \num{40000}  &             &                 &         &             & $\ang{81}$         \\
\citet{Yokuda1990}                         & EXP      & \num{91000}  &             &                 &         &             & $\ang{78}$         \\
Heimenz (see \citet{White1991})            & EXP      &              &             &                 &         &             & $\ang{80.5}$       \\
\citet{Achenbach1968}                      & EXP      & \num{100000} &             &                 &         &             & $\ang{78}$         \\
\citet{Salvatici2003}                      & LES      & \num{20000}  & $0.98-1.17$ & $0.38-0.46$     &         & $0.96-1.12$ &                \\
\citet{Wornom2011}                         & LES      & \num{20000}  & $1.27$      & $0.6$           & $0.19$  & $1.09$      & $\ang{86}$         \\
\citet{Lysenko2014}                        & LES      & \num{20000}  & $1.3$       & $0.75$          & $0.2$   & $1.2$       & $\ang{88}$         \\
\citet{Lysenko2014}                        & LES      & \num{20000}  & $1.32$      & $0.64$          & $0.2$   & $1.05$      & $\ang{86}$         \\
Current (fine mesh)                        & DES      & \num{27000}  & $1.31$      & $0.50$          & $0.215$  & $1.46$      & $\ang{81}$         \\
Current (coarse mesh)                      & DES      & \num{27000}  & $1.19$      & $0.47$          & $0.16$  & $1.27$      & $\ang{79}$         \\ \bottomrule
\end{tabular}
\end{table}

The current results with DES are in agreement with the experimental data and LES simulations; especially with \citet{Norberg1994,Norberg2003} and \citet{West1993} who had values at $\text{Re} = \num{27000}$. $\text{St}$ was underestimated for the coarse mesh because a coarse time step was chosen to reduce the wall clock time of simulations. Seven time steps per shedding cycle were used and $\text{St} = 0.19$ when the time step was reduced; as the oscillations were accurately resolved. The fine mesh case in Table~\ref{tab:DES_summary} is the result of the independence study on both the temporal and spatial discretisations.

\begin{figure} \centering
  \includegraphics{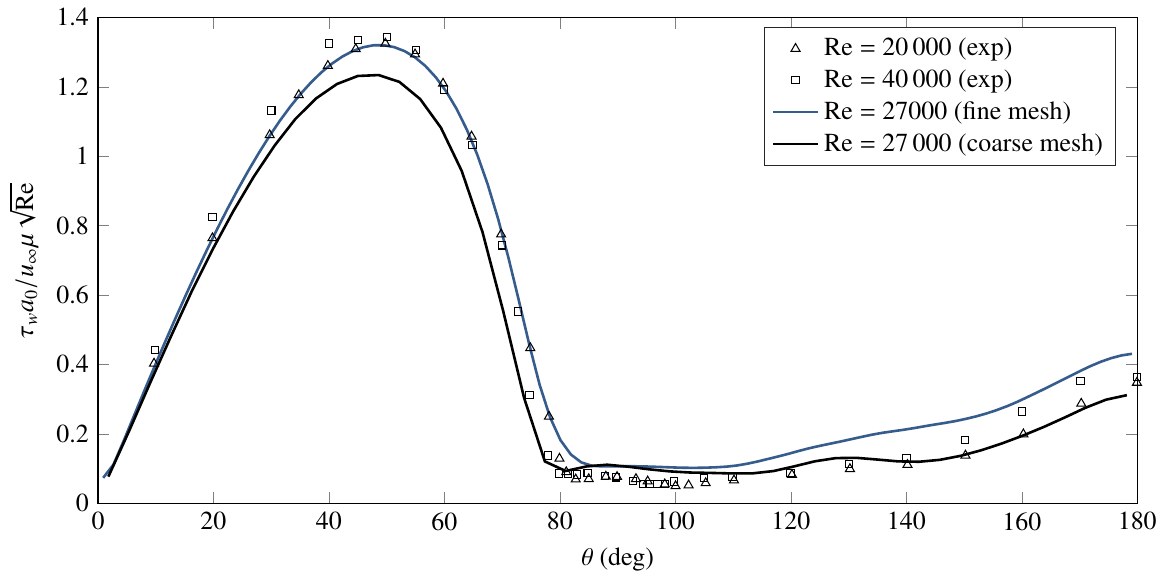}
  \caption{Time-averaged wall shear stress distribution around a rigid circular cylinder using 3-D unsteady DES simulations (symmetric about the stagnation point). The $\tau_w$ has been made dimensionless to compare with the experiments by \citet{Son1969}.
\label{fig:DES_wallshear}}
\end{figure}

Shearing forces on the cylinder wall was the driving mechanism of the receding boundary because the erosion rate was directly proportional to this $\tau_w$ distribution. Therefore an accurate $\tau_w$ profile around the cylinder was the key objective; a favourable comparison with experiments is shown in Figure~\ref{fig:DES_wallshear}. The $\tau_w$ is scaled with $\sqrt{\text{Re}}$ as boundary layer theory predicts this relationship \citep{Son1969} such that the measurements collapse on a single curve. The $\tau_w$ approached zero at the stagnation point and peaked around $\theta \approx \ang{50}$: corresponding to the highest rate of erosion for the initial cylinder.

The separation point $\theta_{\text{sep}} = \ang{79}$ is where $\tau_w$ transitions from a high value upstream to a significantly lower value downstream. Overpredicting $\theta_{\text{sep}}$ gave a more streamlined evolution of the cylinder as this high $\tau_w$ was overestimating the erosion on the sides of the cylinder. The profile of $\tau_w$ in the wake was resolved accurately (Figure~\ref{fig:DES_wallshear}) and allowed simulation of erosion due to the shearing beyond $\theta_{\text{sep}}$ (in contrast to the steady 2-D laminar approximation as shown in Figure~\ref{fig:laminar_wallshear}).

\begin{figure} \centering
  \begin{subfigure}[b]{\textwidth} \centering
    \includegraphics{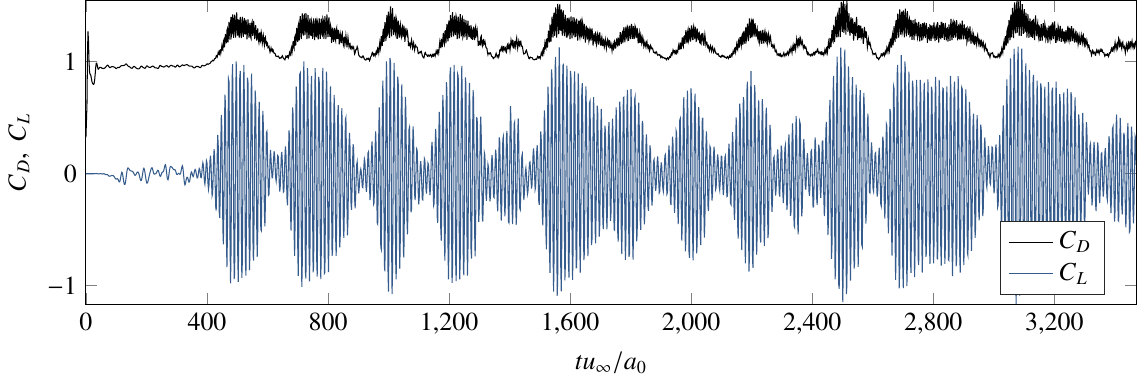}
    \caption{\centering\label{fig:DES_coefficients_smooth}}
  \end{subfigure}%
  \newline\newline
  \begin{subfigure}[b]{\textwidth} \centering
    \includegraphics{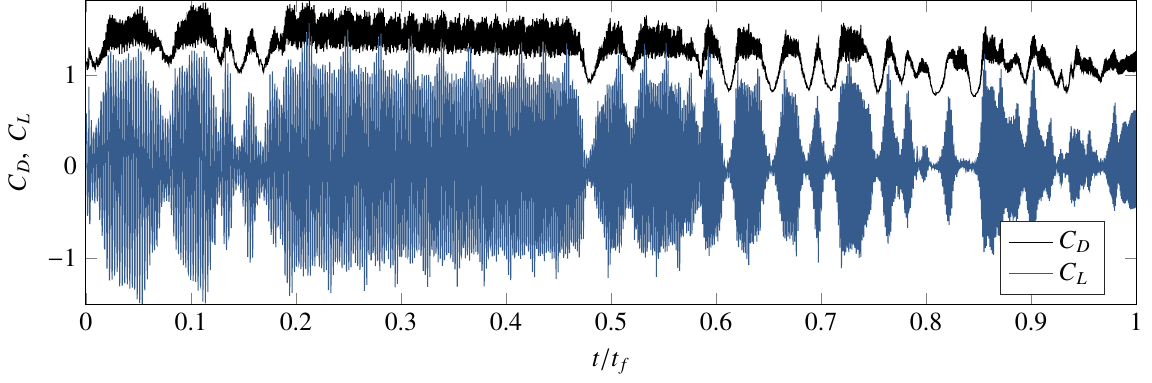}
    \caption{\centering\label{fig:DES_coefficients_MDM}}
  \end{subfigure}
  \caption{Drag and lift coefficients using 3-D unsteady DES simulations for (\subref{fig:DES_coefficients_smooth}) the initial rigid cylinder from initial conditions and (\subref{fig:DES_coefficients_MDM}) the eroding cylinder case starting from the developed flow field. The time range of (\subref{fig:DES_coefficients_smooth}) is approximately $2/3$ that of (\subref{fig:DES_coefficients_MDM}).\label{fig:DES_coefficients}}
\end{figure}

The flow field was first developed for the rigid cylinder case as shown in Figure~\ref{fig:DES_coefficients_smooth}. Transient statistics used in Table~\ref{tab:DES_summary} were sampled from $t \approx \num{600} u_\infty / a_0$ across \num{228} vortex shedding cycles: where the flow was fully developed. The non-uniform amplitudes of $C_D$ and $C_L$ are a feature of vortex shedding in the spanwise direction and are observed both in experiments \citep{Schewe1983} and simulations \citep{Lysenko2014}. This developed solution was then used as the initial conditions for the deforming mesh simulations; across \num{637} vortex shedding cycles. $C_D$ and $C_L$ gradually peak at $t \approx 0.2 t_f$ ($\SI{20}{\percent}$ through the simulation) as the cylinder transforms from its initial shape; with peak erosion occurring $\theta \approx \pm \ang{50}$.  The $C_D$ and $C_L$ then decrease and plateau as the terminal form is reached. The total force acting on the cylinder reduces with time significantly as the projected area against the flow also decreases; however, the coefficients are scaled with $A = A(t)$.

\begin{figure} \centering
  \includegraphics{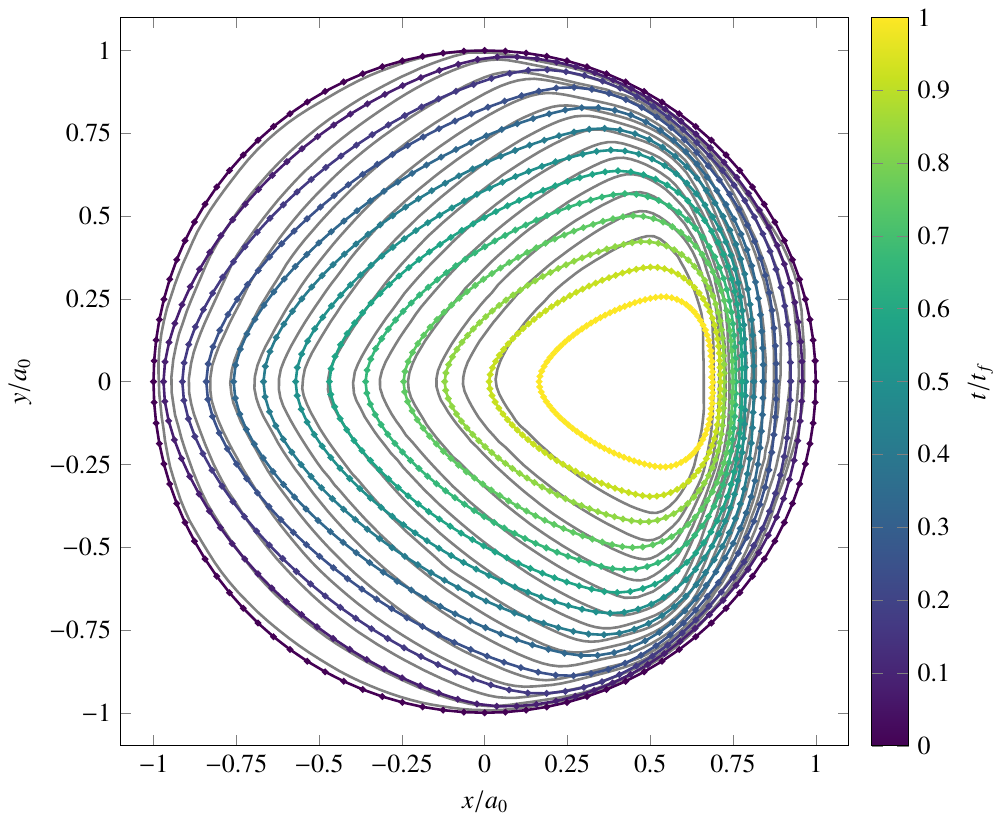}
  \caption{Erosion of a clay cylinder in water (flow is from left to right) using unsteady DES in 3-D (with 2-D shape evolution). The coloured lines show the evolution of the cylinder profile where boundary nodes are represented with points. The erosion factor $\eta = \num{1.4e-3}$ and 600 mesh updates were performed (every 50th is shown here). The solid grey lines are from the experiment \citep{Ristroph2012} where the profiles are on a separate time scale, are linearly spaced in time and terminate before $t_f$ by \SI{8.2}{\percent}.
  \label{fig:DES_profile}}
\end{figure}

The cross sectional area evolution of the cylinder closely followed the same 4/3 power law and experimental values as the steady 2-D laminar case shown in Figure~\ref{fig:laminar_area}. The proportion and profile evolution on the leeward side of the cylinder matched the experiment (Figure~\ref{fig:DES_profile}) indicating that $\theta_\text{sep}$ and the erosion in the wake was accurately captured throughout the cylinder deformation. However, the aspect ratio (spanwise to streamwise $2a/l$) was underestimated where the erosion $v_n$ at the stagnation point was smaller when compared with the proportion of erosion on the sides of the cylinder in the experiment. The curvature ($v_\kappa$) and wall shear stress ($v_\tau$) erosion components were higher on the sides than at the stagnation point but had lower erosion rates. Therefore, another physical mechanism (absent in the model) was responsible for this quick propagation at the leading edge.

\begin{figure} \centering
  \begin{subfigure}[b]{0.5\textwidth} \centering
    \includegraphics{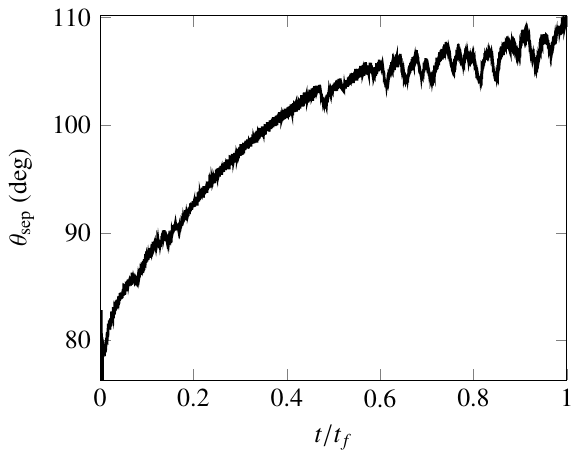}
    \caption{\centering\label{fig:DES_separation_angle}}
  \end{subfigure}%
  \begin{subfigure}[b]{0.5\textwidth} \centering
    \includegraphics{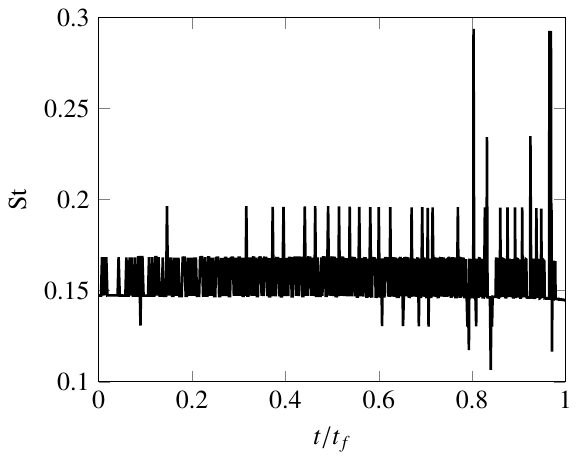}
    \caption{\centering\label{fig:DES_strouhal}}
  \end{subfigure}
  \caption{(\subref{fig:DES_separation_angle}) Separation angle and (\subref{fig:DES_strouhal}) Strouhal number of the evolving single cylinder with unsteady DES in 3-D.\label{fig:DES_info}}
\end{figure}

The separation angle increased from $\theta_\text{sep} = \ang{79}$ as the cylinder transformed from its initial circular shape into a more streamlined rounded triangular form. The separation point tended to $\theta_\text{sep} \approx \ang{106}$ (Figure~\ref{fig:DES_separation_angle}) as the terminal flow was reached. Oscillations of $\Delta \theta = \ang{3}$ were caused by the unsteady nature of the flow.

A constant $\text{St}$ was expected across the Reynolds range experienced by the cylinder as shown in experiments \citep{Son1969}. The $f_v$ increased inversely proportional to $a$ giving a constant $\text{St}$ (Figure~\ref{fig:DES_strouhal}) as predicted from Equation~\ref{eqn:Strouhal_number}. The final maximum radial distance $a_f = 0.25 a_0$ yields a final Reynolds number of $\text{Re}_f = \num{7000}$. Similarly, $\Delta \tilde{t}$ was scaled with $a$ from Equation~\ref{eqn:time_step} giving a uniform resolution of seven time steps per vortex cycle. The discrete jumps of $\text{St}$ in Figure~\ref{fig:DES_strouhal} straddling the mean value were caused by a numerical artefact of this coarse time resolution; vortex periods were measured between peaks of $C_L$ and these discrete steps correspond to $T_v/7$.

\subsection{Sparsely spaced lattice of cylinders}
\begin{figure} \centering
  \includegraphics{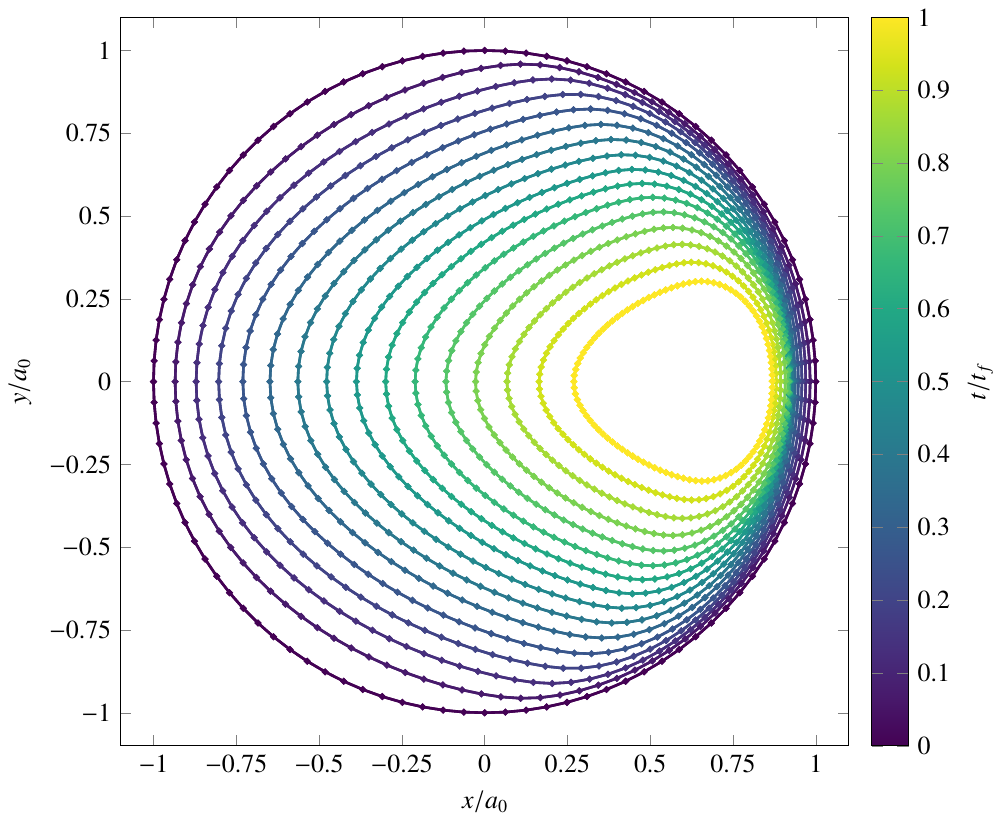}
  \caption{Erosion of a clay cylinder within a lattice of cylinders in water (flow is from left to right) using unsteady DES in 3-D. The coloured lines show the evolution of the cylinder profile where boundary nodes are represented with points. The erosion factor $\eta = \num{1.4e-3}$ and 750 mesh updates were performed (every 50th is shown here).\label{fig:lattice_profile}}
\end{figure}

The cylinder within an infinite lattice of evolving cylinders produced a more rounded shape (Figure~\ref{fig:lattice_profile}) compared to the single cylinder case (Figure~\ref{fig:DES_profile}). The lattice cylinder had a greater portion of erosion on the upstream face than the downstream face because of the turbulent, varied and oscillatory flow in the wake of the upstream cylinders. The dominant influence on the cylinder was from the neighbouring cylinder directly upstream; the flow field and shear profiles are shown as videos in the supplementary material.

\begin{figure} \centering
  \begin{subfigure}[b]{0.5\textwidth} \centering
    \includegraphics{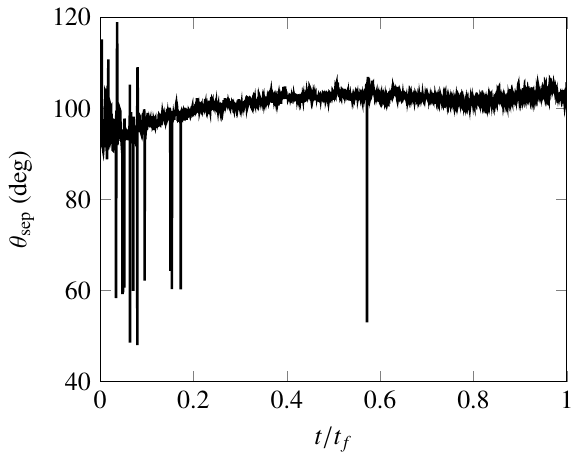}
    \caption{\centering\label{fig:lattice_separation_angle}}
  \end{subfigure}%
  \begin{subfigure}[b]{0.5\textwidth} \centering
    \includegraphics{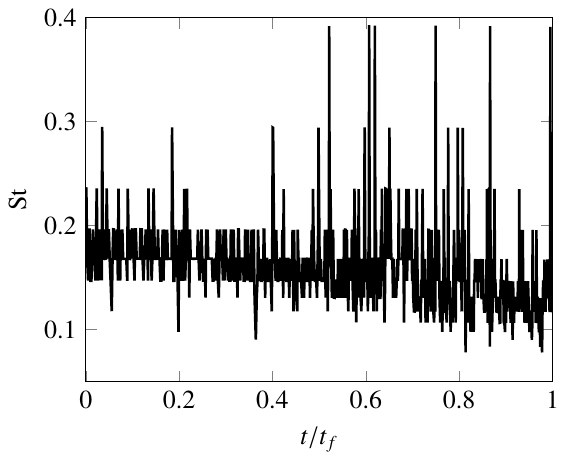}
    \caption{\centering\label{fig:lattice_strouhal}}
  \end{subfigure}
  \caption{(\subref{fig:lattice_separation_angle}) Separation angle and (\subref{fig:lattice_strouhal}) Strouhal number of the evolving cylinder in a sparsely spaced lattice with unsteady DES in 3-D.\label{fig:lattice_info}}
\end{figure}

A terminal form is also attained for the cylinder lattice where the separation angle tends to $\theta_\text{sep} \approx \ang{103}$ (Figure~\ref{fig:lattice_separation_angle}). The Strouhal number appears to decrease slightly in Figure~\ref{fig:lattice_strouhal}. Both $\theta_\text{sep}$ and $\text{St}$ have significant scatter because of the chaotic periodic inlet conditions: the wake of the upstream cylinders.

\subsection{Closely packed lattice of cylinders}
\begin{figure} \centering
  \includegraphics{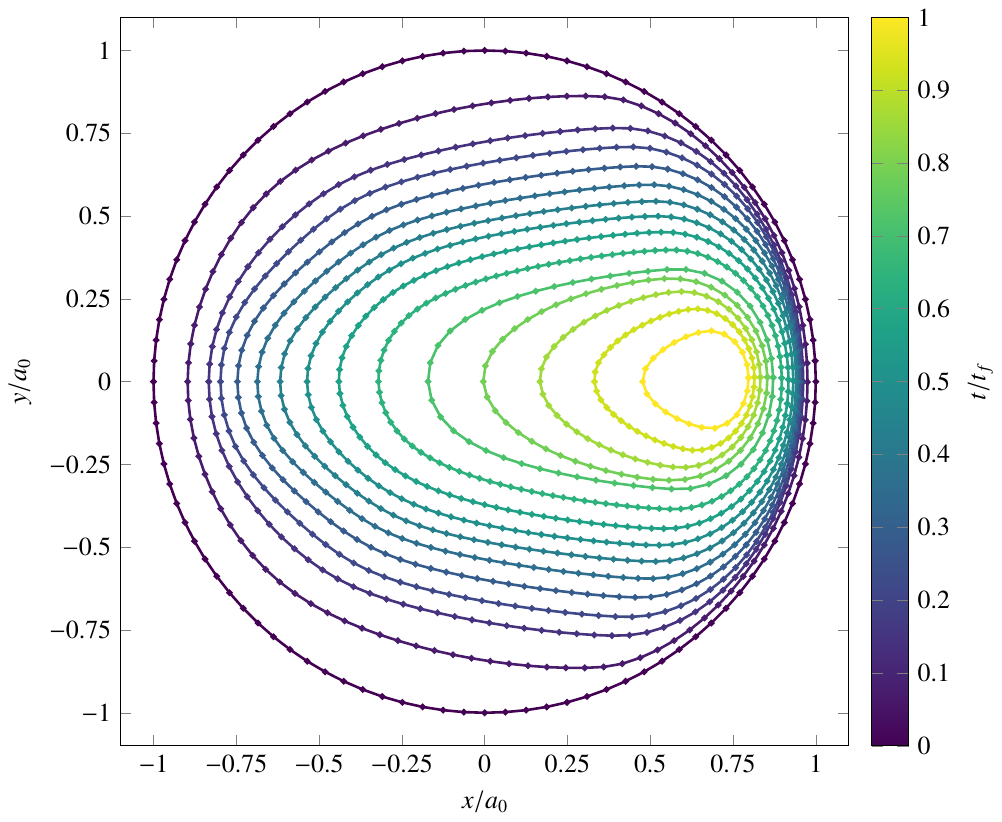}
  \caption{Erosion of a clay cylinder within a closely packed lattice of cylinders in water (flow is from left to right) using unsteady DES in 3-D. The coloured lines show the evolution of the cylinder profile where boundary nodes are represented with points. The erosion factor $\eta = \num{1.4e-3}$ and 734 mesh updates were performed (every 50th is shown here).\label{fig:packed_profile}}
\end{figure}

Flow surrounding the closely packed lattice with $L/l_0 = 2$ transitioned from a reattachment regime to the co-shedding regime (detailed in Section~\ref{sec:lattice_of_cylinders}). Initially, the cylinder experienced a higher rate of erosion on the sides compared with the sparsely spaced cylinder (Figure~\ref{fig:lattice_profile}), as shown in Figure~\ref{fig:packed_profile}, as a result of the shear layer reattachment between cylinders. The leading edge receded at a slower rate than the sides, forming a more streamlined profile. However, the gap dynamics provided transient $\tau_w$ on this leading edge and a higher rate was simulated when compared to the laminar upstream conditions (Figure~\ref{fig:DES_profile}).

\begin{figure} \centering
  \begin{subfigure}[b]{0.5\textwidth} \centering
    \includegraphics{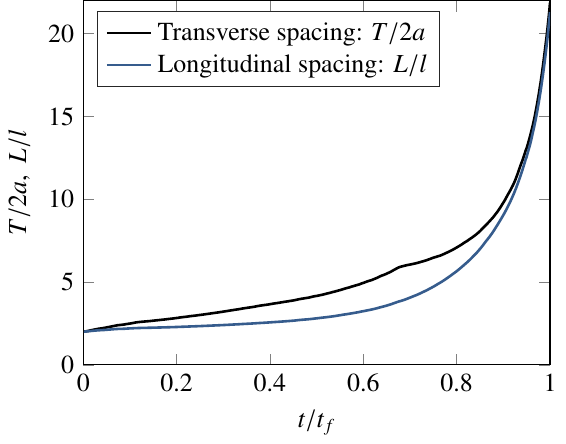}
    \caption{\centering\label{fig:packed_spacing}}
  \end{subfigure}%
  \begin{subfigure}[b]{0.5\textwidth} \centering
    \includegraphics{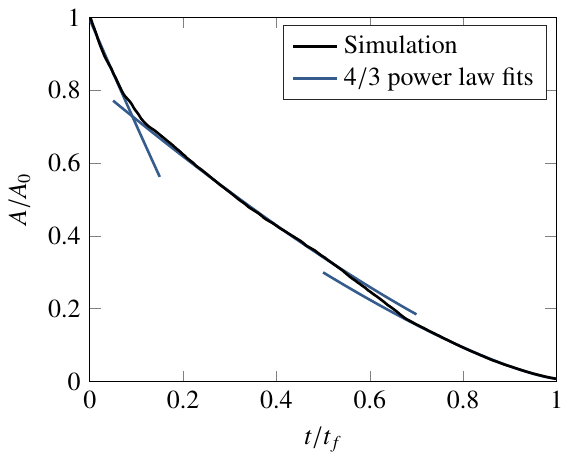}
    \caption{\centering\label{fig:packed_area}}
  \end{subfigure}
  \caption{Closely packed lattice of cylinders using unsteady DES in 3-D (\subref{fig:packed_spacing}) Transverse (normal to flow) and longitudinal (parallel with flow) spacings between cylinders in the closely packed lattice, normalised with their respective cylinder lengths $2a$ and $l$. (\subref{fig:packed_area}) Cross sectional area evolution normalised by the initial area.\label{fig:packed_info}}
\end{figure}

Pitch ratio $L/l$ consistently increased as the body eroded: $L$ remained constant (cylinders were fixed) whereas $l$ decreased. Aspect ratio $2a/l$ initially reduced to a minimum of $0.65$ at $t/t_f = 0.64$ before increasing to a terminal ratio of approximately unity, as shown in Figure~\ref{fig:packed_spacing}. This inflection point offers a suggestion for the transition between the two flow regimes at $L/l = 3.5$ and agrees with experiments \citep{Zdravkovich1987} which observed this regime change at $L/l = 3.4-3.8$. The cylinder then erodes in a similar form as the sparsely spaced cylinder (entirely within the co-shedding flow regime) where the body tends to a terminal form and experiences self-similar evolution.

The kink at $t/t_f = 0.68$ in Figure~\ref{fig:packed_spacing} (and the slower erosion rate in Figure~\ref{fig:packed_profile}) is non-physical, but a numerical artefact that corresponds to the manual remesh where the surrounding grid resolution increases. The number of cells are conserved for each mesh update, causing a lower grid resolution over time as the cylinder shrinks and the fluid volume increases. The kink was not apparent for $L/l$ because the decrease of $l$ was smaller than for $2a$; less of a change in grid resolution. However, the inflection point was slightly before this time and the body had begun to transition to the co-shedding regime beyond the remeshing point.

Figure~\ref{fig:packed_profile} shows a non-uniform rate of erosion through time; the evolution of $\tau_w$ is shown with videos in the supplementary material. The leading edge initially erodes quickly until slowing at $t/t_f \approx 0.2$ (where the spacing between the cylinder profiles are smallest) before eroding quickly again. The time of slowest simulated erosion had a pitch ratio of $L/l = 2.3$ (Figure~\ref{fig:packed_spacing}) and this spacing is in agreement with the lowest observed drag in experiments \citep{Zdravkovich1977}.

Both relative spacings $L/l$ and $T/2a$ increased as shown in Figure~\ref{fig:packed_spacing}. The aspect ratio tended to a lower value for the reattachment regime and then returned near unity within the co-shedding flow regime. $\text{Re}$ scales linearly with $a$ and $\text{Re}_f = \num{2500}$ (still within the subcritical flow range).

Cross sectional area (Figure~\ref{fig:packed_area}) also decreases following a $4/3$ power law for this closely packed cylinder case, split into three sections for covering the different flow regimes. The slope of $A(t)$ is steeper in the reattachment regime compared to the co-shedding flow regime. The power law has a closer fit for the terminal form as $t \to t_f$ with consistent vortices shedding from upstream, and is a rougher fit for the transitioning period where the eddies between cylinders have a wide range of dynamics \citep{Sumner2010}.

\section{Discussion}
\subsection{Flow patterns}
Laminar flow upstream conditions were applied for both single cylinder cases using the laminar and DES viscous models. $\text{Re} = \num{27000}$ is within the subcritical flow regime and the shear layers from the sides of the cylinder formed K\'arm\'an vortices in the wake. The steady laminar approximation neglected $\tau_w$ in the wake beyond $\theta_\text{sep}$, and was in good agreement with theory \citep{Moore2013}: the body evolved into a terminal form with a right angle pointed into the direction of flow (Figure~\ref{fig:laminar_profile}) with near uniform wall shear stress (Figure~\ref{fig:laminar_wallshear}). Modelling the vortex structures in the wake of the cylinder with scale resolving simulations allowed erosion based on $\tau_w$ on the leeward face. The location of lowest $\tau_w$, and thus erosion rate, was accurately captured on the sides of the cylinder (Figure~\ref{fig:DES_profile}) and the proportion of erosion of the windward and leeward faces match reasonably well with experiment \citep{Ristroph2012}.

For the lattice arrangement cases, the cylinders block a portion of the fluid volume causing significant blockage. The blockage ratio $2a/T$ is defined as the ratio of blocked (cylinder diameter $2a$) to total (transverse spacing $T$) cross sectional length normal to the flow and cylinder axis. Our initial blockage was $2a_0/T = \SI{25}{\percent}$ for the sparsely spaced case and $2a_0/T = \SI{50}{\percent}$ for the closely packed case. Blockage as low as $\SI{16}{\percent}$ causes an increase in drag as observed in experiments at $\text{Re} = \num{30000}$ \citep{West1982}. Consider an effective Reynolds number
\begin{equation}
\text{Re}_\text{eff} = \frac{2 u_\infty a}{\nu} \frac{T}{T-2a}
\end{equation}
where the second term on the right hand side scales $u_\infty$ to the local velocity between cylinders (from the conservation of mass principle). This term tends to unity for $T \to \infty$ (unblocked flow) and tends to infinity for $T \to 2a$ (fully blocked lattice with no flow). $\text{Re}_\text{eff}$ gives a rough approximation to compare flow properties between our cases and of cylinder arrays in literature without blockage; particularly the cases where there is minimal proximity interference but wake interference is important (for comparing with tandem arrangements).

Erosion rates of cylinders in higher speed flow are expected to be greater since $\tau_w \propto \sqrt{\text{Re}}$ as discussed by \citet{Son1969} and their shear distributions are shown in Figure~\ref{fig:DES_wallshear}. This correlation was observed between the three cases using unsteady DES. The closely packed lattice with $\text{Re}_\text{eff,0} = \num{55000}$ had the quickest initial erosion rates, followed by the sparsely spaced lattice with $\text{Re}_\text{eff,0} = \num{37000}$ and then the single cylinder with $\text{Re}_\text{eff} = \num{27000}$. Both of these $\text{Re}_\text{eff,0}$, and consequently all $\text{Re}_\text{eff}$, remain in the subcritical flow regime and therefore have similar flow patterns throughout the cylinder evolution process.

The speed of vortex shedding increased as the cylinder eroded with $f_{v,f}/f_{v,0} \approx 4$ and $a_f/a_0 \approx 4$ yielding a constant $\text{St}$ as shown in Figure~\ref{fig:DES_strouhal} and predicted with Equation~\ref{eqn:Strouhal_number}. This relatively constant Strouhal number supports the notion that K\'arm\'an vortices scale with the characteristic length normal to flow, $2a$, across an arbitrary bluff body. The simulation time step $\Delta \tilde{t}$ was scaled with $a$ (Equation~\ref{eqn:time_step}) allowing the simulation to capture the vortex shedding from slower $f_{v,0}$ to faster $f_{v,0}$ speeds throughout the cylinder evolution.

\subsection{Terminal form}
\begin{figure} \centering
  \begin{subfigure}[b]{0.48\textwidth} \centering
    \includegraphics{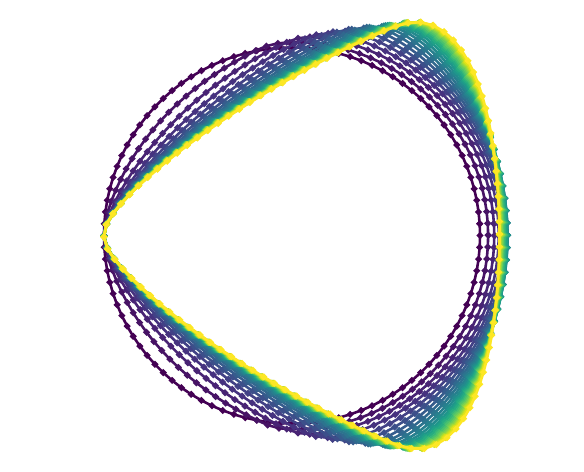}
    \caption{\centering\label{fig:laminar_profile_scaled}}
  \end{subfigure}%
  \begin{subfigure}[b]{0.48\textwidth} \centering
    \includegraphics{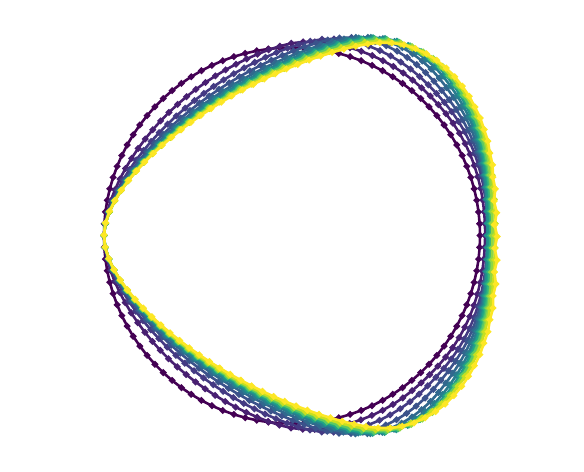}
    \caption{\centering\label{fig:DES_profile_scaled}}
  \end{subfigure}
  
  \begin{subfigure}[b]{0.48\textwidth} \centering
    \includegraphics{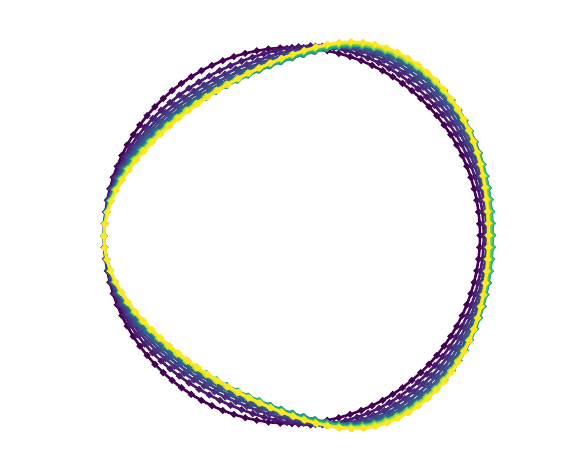}
    \caption{\centering\label{fig:lattice_profile_scaled}}
  \end{subfigure}%
  \begin{subfigure}[b]{0.48\textwidth} \centering
    \includegraphics{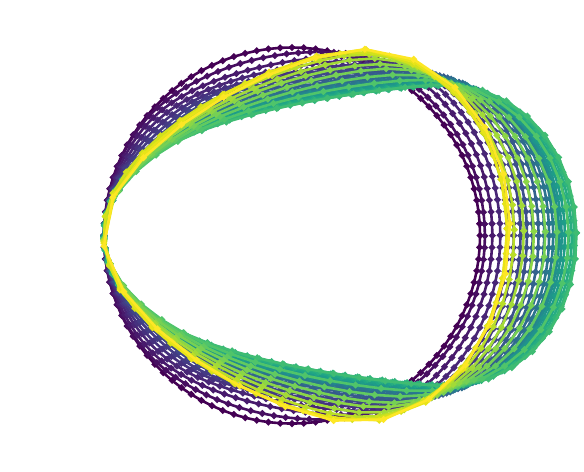}
    \caption{\centering\label{fig:packed_profile_scaled}}
  \end{subfigure}
  \caption{Scaled profiles of the four cases: (\subref{fig:laminar_profile_scaled}) single cylinder, 2-D steady laminar with $\epsilon = 0.1$ (\subref{fig:DES_profile_scaled}) single cylinder, 3-D unsteady DES with $\epsilon = 0.4$ (\subref{fig:lattice_profile_scaled}) sparsely spaced cylinder lattice, 3-D unsteady DES with $\epsilon = 0.4$ (\subref{fig:packed_profile_scaled}) closely packed lattice, 3-D unsteady DES with $\epsilon = 0.4$ (doubled the number of mesh updates shown here for clarity of evolution). Colours are as per their respective profiles in Figures \ref{fig:laminar_profile}, \ref{fig:DES_profile}, \ref{fig:lattice_profile} and \ref{fig:packed_profile}.\label{fig:scaled_profiles}}
\end{figure}

In all four cases, the cylinder eroded towards a terminal form and then continued eroding self-similarly as shown in Figure~\ref{fig:scaled_profiles} with an aspect ratio of approximately unity, regardless of laminar or unsteady turbulent upstream conditions. The steady laminar approximation case (Figure~\ref{fig:laminar_profile_scaled}) took longer to reach its terminal form compared to the unsteady DES case (Figure~\ref{fig:DES_profile_scaled}), specifically around the sides, because the evolving profile tracked around the leeward face (no erosion downstream).

The terminal profile of the sparsely spaced lattice (Figure~\ref{fig:lattice_profile_scaled}) was reached quicker than the single cylinder cases. Oscillating vortices were shed from the preceding cylinder (modelled with a periodic boundary) and provided a consistent upstream flow field throughout the evolution. The subcritical flow regime remained from start to finish and the relative size and frequency of the vortices remained roughly constant with the cylinder ($\text{St}$ in Figure~\ref{fig:lattice_strouhal}). Whereas for the single cylinder arrangements, the fluid structure coupling had a time dependence where the streamlines followed the deforming boundary, altering the $\tau_w$ distribution (Figure~\ref{fig:laminar_wallshear}), subsequently delaying the evolution to its terminal form. This more gradual profile development was also observed in the experiment \citep{Ristroph2012}.

Initially, the cylinder profile in the closely packed lattice tended to a more streamlined body (Figure~\ref{fig:packed_profile_scaled}) as $\tau_w$ was higher on the sides where the shear layer reattached from the cylinder upstream. The flow transitioned from the reattachment regime to the co-shedding regime as the relative spacing between cylinders increased. The terminal form of this configuration was similar to the sparsely spaced cylinders because both cases had similar wake interference characteristics.

\subsection{Erosion rates}
Erosion at the leading edge was naturally driven by $\tau_w$ in the lattice of cylinders (Figure~\ref{fig:lattice_profile}), whereas the leading edge was driven by the curvature correction for the single cylinder (erosion begins after a curvature is formed, as shown in Figure~\ref{fig:DES_profile}). The closely packed lattice featured several erosion rates (spacings between contours in Figure~\ref{fig:packed_profile}) as the flow transitioned from the reattachment regime to the co-shedding regime.

Cross sectional area of all eroding cylinders followed a $4/3$ power law in time for a wide range of Reynolds number in the subcritical flow regime (Figures~\ref{fig:laminar_area} and \ref{fig:packed_area}) as predicted by theory (Equation~\ref{eqn:power_law}). This relationship was observed in both experiment \citep{Ristroph2012} and simulations with consistent upstream conditions (Figure~\ref{fig:laminar_area}). However, the upstream flow conditions in the closely packed lattice changed significantly as the cylinder eroded and the flow regimes transitioned. A steeper slope of the $4/3$ power law fit was used in the heavily blocked flow period $t/t_f < 0.1$ compared to when the flow had negligible blockage effects at $t/t_f > 0.7$. This discrepancy can be explained by the difference in local velocity between the cylinders where $\text{Re}_\text{eff} / \text{Re} > 1$ had higher $\tau_w$ and thus erosion rate when compared to the low blocked case with $\text{Re}_\text{eff} \approx \text{Re}$. The transition time between these periods had vastly changing gap dynamics but roughly followed the power law relationship (Figure~\ref{fig:packed_area}).

\subsection{Opening angle}
{The opening angle $\psi$ at the nose of the cylinder (pointed upstream) has been calculated by fitting a circle to each of the top and bottom sides of the cylinder and finding their angle of intersection, as shown in Figure~\ref{fig:nose_diagram}. Nodes were selected on the top and bottom sides according to their normal vector components, with a tolerance to exclude nodes near the stagnation point where the curvature was significantly different. The opening angle $\psi$ of all the eroding cylinder cases begin with $\psi = \ang{180}$ (concentric circles of the top and bottom sides) and plateau to a final value as the cylinder morphs into its final shape and erodes self-similarly.

Exact solutions predict a right angled wedge ($\psi = \ang{90}$) would produce uniform wall shear stress on the body \citep{Moore2013}. The Prandtl-based model closely matched this analytical prediction with a final $\psi = \ang{91}$. The 2-D steady laminar simulation tended towards $\psi = \ang{95}$ for its self-similar evolution (Figure~\ref{fig:nose_angle}), indicating that the laminar approximation roughly matches the right angled wedge shape.

In contrast, the experiment had a wider nose angle of $\psi \approx \ang{120}$ for its final form as shown in Figure~\ref{fig:nose_angle}. The 3-D unsteady DES simulation captured the transition from the viscous layer to the developed shear layer causing oscillating vortical structures and transient effects along the span of the cylinder; yielding a $\psi = \ang{118}$ which matches well with the experiment.

\begin{figure} \centering
  \begin{subfigure}[b]{0.5\textwidth} \centering
    \includegraphics{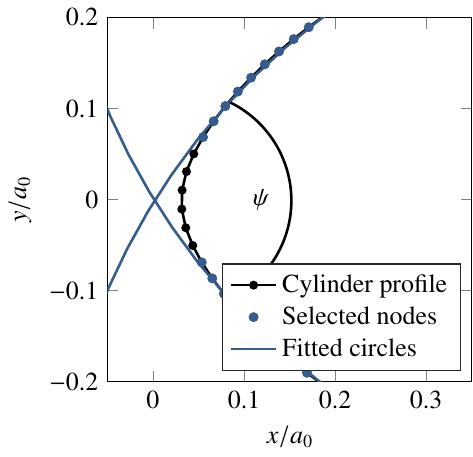}
    \caption{\centering\label{fig:nose_diagram}}
  \end{subfigure}%
  \begin{subfigure}[b]{0.5\textwidth} \centering
    \includegraphics{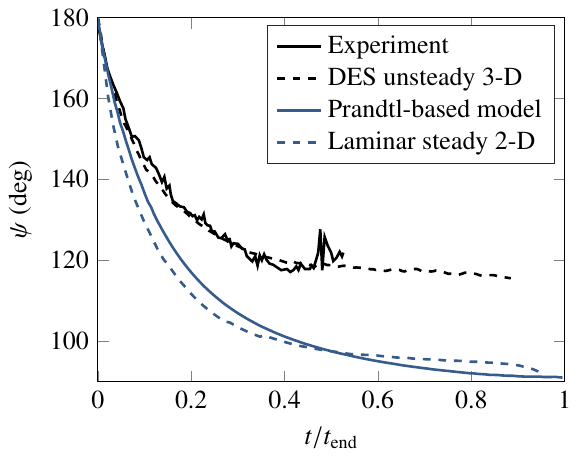}
    \caption{\centering\label{fig:nose_angle}}
  \end{subfigure}
  \caption{(\subref{fig:nose_diagram}) Opening angle $\psi$ of the deforming cylinder boundary partway through a simulation. (\subref{fig:nose_angle}) Opening angles over time for the deforming cylinder cases compared with experimental \citep{Ristroph2012} and theoretical measurements \citep{Moore2013}.\label{fig:nose_info}}
\end{figure}

\subsection{Curvature component} \label{sec:curvature_component}
A curvature dependence was introduced for the interface evolution in Equation~\ref{eqn:erosion_velocity} as suggested by \citet{Moore2013}. They used this term for ensuring regularity of their numerical methods but also found that smoothing the sharp edges replicated certain features in the experiment by \citet{Ristroph2012}; including the roundedness of the leading edge and around the regions of separation on the sides of the terminal body. Flow at the stagnation point was stationary giving $\tau_w = \SI{0}{\pascal}$, as shown in Figure~\ref{fig:laminar_wallshear} (face values are plotted although the stagnation point sits on a cell node), which would not erode without a curvature term; causing a sharp point to form at the stagnation point and ultimately leading to a singularity.

The curvature term appears to model a real physical mechanism experienced by the clay cylinder in the experiment which would otherwise be absent in our simulations. Two possible sources of this mechanism are small scale flow fluctuations and the slow dissolution of clay in water \citep{Moore2013}. Another potential factor for erosion could be the static pressure. \citet{Mercier2014} modelled a submerged turbulent impinging jet and found the static pressure at the stagnation point increased as the soil eroded; this static pressure appears to increase by a similar factor that their peak $\tau_w$ decreased.

Erosion from this other mechanism (whether its explicitly or implicitly curvature dependent) seems to follow the $\sqrt{L/L_0}$ factor used in Equation~\ref{eqn:erosion_velocity} and suggested by \citet{Moore2013}. This erosion term $v_\kappa$ may be a non-linear addition to erosion by wall shear stress $v_\tau$ and possibly describes a lower limit for erosion such that $v_n = \max{(v_\kappa,v_\tau)}$. This hypothesis is based on observing the curvature evolution for the quasi triangular terminal shape for the single cylinder case (Figure~\ref{fig:DES_profile}). Our simulations with $\epsilon = 0.4$ matched the experimental profile at the nose (zero $\tau_w$) but overpredicted the curvature at $\theta_\text{sep}$ (non-zero average $\tau_w$ from the fluctuating flow field).

\subsection{Future work}
There were some differences between the simulations and experiment, specifically the aspect ratio of the terminal form was underestimated as shown in Figure~\ref{fig:DES_profile}. However, our cylinder profiles were generally in agreement with the experiment \citep{Ristroph2012} for the proportion of erosion in the wake and on the sides. The curvature component in Equation~\ref{eqn:erosion_velocity} and discussed in Section~\ref{sec:curvature_component} appears to provide an approximate correction although this term could be explored further.

One limitation in the simulations was the coarse grid and time resolutions due to the computational restrictions inherent in scale resolving simulations. Future work could include resolving the vortices and particularly $\theta_\text{sep}$ by refining the mesh and increasing the number of time steps per vortex period. Other limitations include the prescribed spanwise extent, uncertainties in turbulence model and uncertainties in experimental measurements.

Experiments with either a different $\text{Re}$ (within the subcritical flow regime) or cylinder arrangement using the same clay material would allow a quantitative validation of erosion rates (compared with our non-dimensional analysis). The material dependent coefficient (absorbed in $\eta$ alongside the pseudo erosion time step in Equation~\ref{eqn:displace_nodes}) could be compared between both cases and in theory should be equivalent.

Other simulations could be performed on tandem cylinders without reattachment using a small pitch ratio of $1 < L/l < 1.2-1.8$. This case would require either a dense mesh within the gaps or added cells throughout the simulation, \textit{remeshing}, to ensure a sufficient grid density to capture the gap dynamics. Another potential setup includes staggered cylinder arrangements which are often used in heat exchangers and have a finite array of cylinders; voiding the assumption of periodic boundaries.

\section{Conclusion}
A cylinder was modelled in high speed flow and compared with existing experiments. The eroded cylinder converged to a rounded triangular shape and continued eroding self-similarly with this terminal form. The upwind face had a uniform shear stress distribution and featured a near right-angled wedge directed into the flow as predicted by theory \citep{Moore2013}; this uniform erosion rate lead to a bluff body rather than a streamlined profile. A cylinder was then simulated within an infinite array of other cylinders, forming a lattice. This arrangement provided an oscillatory motion of vortices upstream and we found a similar terminal form emerged. The evolving profile of the cylinder was sensitive to the surrounding flow field; a closely packed lattice resulted in an intermediate profile more streamlined than its terminal form when the flow field closely resembled the sparsely separated case. However, a similar ultimate terminal form emerged regardless of the initial and intermediate shapes. Numerical simulations are a powerful tool for predicting and exploring physical phenomena and we have shown that CFD can accurately capture the fluid structure interaction of an eroding cylinder with large shape and size deformation.

\section*{References}

\bibliography{CylinderErosion}

\end{document}